\documentstyle[prl,twocolumn,aps]{revtex}
\begin{document}
\draft

\title{\centerline{Effect of spin degeneracy on scaling}
       \centerline{in the quantum Hall regime}}
\author{C.~B.~Hanna$^{(1)}$, D.~P.~Arovas$^{(2)}$,
K.~Mullen$^{(1,a)}$, and S.~M.~Girvin$^{(1)}$}
\address{
$^{(1)}$Department of Physics, Indiana University,
Bloomington, Indiana 47405\\
$^{(2)}$Department of Physics, U.~C. San Diego,
La Jolla, California 92093\\ }

\maketitle

\begin{abstract}
\baselineskip=14pt
We study the delocalization transition
for spin-polarized and spin-degenerate non-interacting electrons
in the lowest Landau level.
We perform finite-size scaling calculations of the Thouless number,
for varying amounts of potential and spin-orbit scattering.
{}For spin-polarized electrons,
we obtain a one-parameter scaling function for the Thouless number
that fits scaled experimental data for the longitudinal resistivity.
{}For spin-degenerate electrons with spin-orbit scattering,
the Thouless number is peaked away from the band center by an amount
proportional to the strength of the spin-orbit scattering.
The universality class of the delocalization transition
for non-interacting spin-degenerate electrons
in the quantum Hall regime
is found to be the same as for spin-polarized electrons.
We also study the density of states and Thouless number
for the model of pure spin-orbit scattering studied by
Hikami, Shirai, and Wegner [Nucl. Phys. {\bf B408}, 415 (1993)],
which represents a different universality class.

\end{abstract}

\narrowtext

{\bf 1. Introduction}

In the quantum Hall effect\cite{prange},
changes in the quantized value of the Hall conductivity $\sigma_{xy}$
as the magnetic field or particle density are varied are due to a
series of metal-insulator transitions\cite{levine,laugh,halp}.
The insulating phase occurs when the chemical potential lies
in a region of localized states.
In the insulating regime, $\sigma_{xy}$ is quantized
to an integer multiple of $e^2/h$ to a very high degree of accuracy
over a finite range of field values,
and the longitudinal conductivity $\sigma_{xx}$ vanishes\cite{vonk}.
Near the critical points, $\sigma_{xx}$ is nonzero and
$\sigma_{xy}$ no longer remains constant.
The widths of the quasi-metallic regimes become smaller as the
temperature is reduced\cite{wei1}.
A theoretical phase diagram\cite{khem3}
for the integer quantum Hall effect (IQHE)
has recently been proposed\cite{kivel} for the transition between
insulating and quantum Hall regimes.

Metal-insulator transitions are usually described using
scaling theories.
In the absence of a magnetic field,
scaling arguments indicate that in two dimensions,
all states are localized in the thermodynamic limit,
leading to an insulating state\cite{abra}.
In three dimensions, there can exist a range of chemical potential
for which the system is a metal, separated by a mobility edge from
an insulating state.
The situation is qualitatively different in two dimensions
in the presence of a strong perpendicular magnetic field.
Semiclassically, the trajectory of a
charged particle in a magnetic field
can be separated into a drift motion along equipotential contours,
together with small gyrations about the drift motion.
In the limit of smoothly varying disorder,
there is a single energy at which
equipotential contours percolate through
a system.
This corresponds to an extended state\cite{iord,trug}.
Scaling theories of the metal-insulator transition in the
quantum Hall regime have been developed,
consistent with the idea that conductivity
depends on disorder and chemical potential
\cite{{levine},{khem1}}.
{}For the integer quantum Hall effect (IQHE),
random potential scattering broadens the Landau-level peaks
in the density of states.
However, it is known that the regions of extended states
are not broadened in energy.
There is a discrete set of energies $\lbrace E_c \rbrace$,
corresponding (for weak disorder)
to the Landau level energies $(n + {1\over2})\hbar \omega_c$,
at which the localization length $\xi$ diverges as
$\xi \sim |E-E_c|^{-\nu}$\cite{pruis}.

The localization exponent $\nu$
has been measured directly in experiments
by varying the width of Hall bars and determining the scaling
behavior of the longitudinal resistivity
$\rho_{xx}$ in the metallic regime,
at very low temperatures\cite{koch}.
It was found that $\nu \approx 2.3$.
The scaling behavior of the IQHE delocalization transition has also
been measured by studying the resistivity
tensor as a function of temperature\cite{wei1}.
The reciprocal of the width $\Delta B$
of the transition region of $\rho_{xx}$
between Hall plateaux,
and the maximum slope of the Hall resistivity
in the transition region, $d \rho_{xy}/dB$,
both diverge with the same exponent $\kappa$:
\begin{equation}
{1\over{\Delta B}} , \left( {{d \rho_{xy}}\over {dB}} \right)_{max}
\propto \; T^{-\kappa} .
\end{equation}
The scaling exponent $\kappa$ is related to the localization exponent
$\nu$ by a third exponent $p$, according to
$\kappa=p/2\nu$\cite{pruis}.
The exponent $p$ describes how the inelastic
scattering length $l_{in} \propto T^{-p}$,
which determines the effective sample size\cite{thou3},
diverges as a function of temperature.
{}For spin-polarized electrons, experiments yield $\kappa \approx 0.42$
\cite{wei1}.
When combined with the experimentally obtained
value $\nu \approx 2.3$ from
Ref.\cite{koch}, this implies $p \approx 2$.
The value of $p$ has been measured by independent means on the same
sample in Ref.\cite{wei3}, and $p \approx 2$ is also found.

It should be noted that
there are other scenarios for explaining the value of $\kappa$.
{}From the point of view of dynamical scaling, $\kappa=1/z\nu$,
where $z$ is the dynamical scaling exponent\cite{{fisher},{sondhi}}.
The value $z\approx 1$ is
consistent with Coulomb interactions controlling the
dynamics\cite{fisher,poly}.
Polyakov and Shklovskii\cite{poly} have argued that hopping transport
is responsible for the broadening of the $\rho_{xx}$ peaks.
Sondhi and Kivelson\cite{sondhi} equate temperature scaling
to (imaginary) time or frequency scaling,
and propose that the result
$\kappa \approx 1/\nu$ is due to the scaling behavior of the
{\it time} scale (rather than the length scale $l_{in}$).
Recently, the microwave frequency dependence of $\sigma_{xx}$
in the IQHE was measured, and the width $\Delta B$ of
$\sigma_{xx}$ was found to scale with the frequency $\omega$
like $\Delta B\propto \omega^\gamma$,
where $\gamma=0.41\pm 0.04$ for spin-split peaks\cite{engel}.
In this work, we explicitly perform
{\it length} rescaling to obtain $\nu$,
and do not directly calculate the
temperature scaling of the conductivity.

It has also been demonstrated experimentally
that higher derivatives of $\rho_{xy}$ scale
according to
\begin{equation}
\left( {{d^n \rho_{xy}}\over{d B^n}}
\right)_{B_c} \propto \; T^{-n\kappa} ,
\end{equation}
where $B_c$ denotes the center of transition region
(where $\rho_{xx}$ has a local maximum), and $n=1,2,3$\cite{wei2}.
Assuming that Eq.~(2) also holds for all higher derivatives,
it follows that in the metallic region, the resistivity
$\Delta \rho_{xy} \equiv \rho(B_c) - \rho(B)$ near $B_c$
is a scaling function of a dimensionless length scale, $\xi/l_{in}$;
{\it i.e.}, of the ratio of the localization
length to the effective system size\cite{wei2}.

A substantial body of numerical work using finite-size
scaling\cite{ando,aoki1,ono,aoki2}
shows that, for non-interacting spin-polarized
electrons, $\nu \approx 2.3$
\cite{{ando},{huck},{huo},{liu},{wang}},
in agreement with experiment\cite{koch} and with
(non-rigorous) theoretical arguments that $\nu=7/3$
\cite{milni}.
However, for experiments with lower
mobility samples in smaller magnetic
fields, the disorder broadening can exceed the Zeeman splitting.
In this case there can occur a
spin-unresolved quantum Hall transition,
with the Hall conductance changing by $\Delta\sigma_{xy}=2e^2/h$
between plateaux.
{}For such effectively spin-degenerate electrons, experiments find that
$\kappa=0.21$, which is half the size of the spin-polarized exponent
\cite{{wei2},{hwang}}.
In addition, the exponent of the frequency dependence
of the width $\Delta B$ of $\sigma_{xx}$ is also halved for
spin-degenerate electrons: $\Delta B\propto \omega^\gamma$
with $\gamma=0.20\pm 0.05$\cite{engel}.
There is experimental evidence that $p$ is the same for
spin-polarized and spin-degenerate electrons in the IQHE,
which, taken at face value, suggests
that $\nu$ doubles in the spin-degenerate case\cite{wei3}.
A similar approximate doubling of
the exponent is seen in network model
simulations when the localization
energy is assumed to diverge at only one
energy\cite{leedkk}.

However, Khmelnitskii has argued that for the case of two overlapped
spin subbands, the extended states should split\cite{khem2},
and indeed network model simulations
of spin-degenerate electrons are fit
somewhat better by assuming that there
is such a splitting\cite{leedkk}.
Polyakov and Shklovskii\cite{poly} have conjectured that the
for an extended-state whose energy
is split due to the SO interaction
by an amount $2E_c \ll \Gamma$,
where $\Gamma$ is the disorder broadening
of the Landau levels,
the localization length has the form
\begin{equation}
\xi \propto \left |{{\Gamma^2}\over{E^2 - E_c^2}}\right|^\nu ,
\end{equation}
and that one recovers the usual value for $\nu$ only very close to
the spin-split energies $\pm E_c$, at sufficiently low temperature.
Otherwise, an apparent doubling of $\nu$ is obtained.
Reference \cite{wang} has argued that in the limit of very smooth
disorder, spin-degenerate electrons diverge with the {\it same} value
of the localization exponent, $\nu \approx 2.3$,
but at two separate energies.
According to these arguments,
the erstwhile doubling of $\nu$ seen in experiments
is due to assuming that the localization length diverges at
a single energy rather than at two nearby energies.

We shall argue from direct numerical calculations that,
in agreement with theoretical arguments\cite{wang}
and network model studies\cite{chalk},
the localization length diverges at two energies $\pm E_c$,
that the localization exponent is $\nu \approx 2.3$,
and that the value of the universal
peak conductivity for $\sigma_{xx}$
is the same as in the spin-polarized case.
Our calculation of the Thouless number is done within a semirealistic
microscopic model, and provides an alternate confirmation of
the energy splitting produced by spin-orbit scattering.
We have also investigated the pure spin-orbit scattering model
of Hikami, Shirai, and Wegner\cite{hsw} which is believed to be
in a different universality class.

{\bf 2. Computational method}

We have calculated the Thouless number for
electrons moving in the continuum, restricted to
the lowest Landau level (LLL).
In this calculation, a noninteracting two-dimensional electron
gas (2DEG) is subject to a perpendicular magnetic field
${\bf B} = {\bbox\nabla\times} {\bf A}$,
where ${\bf A}$ is the vector potential.
The results presented here were obtained from finite-size
calculations in the LLL, using a basis of continuum states.
{}For a continuum system without disorder
in the presence of total flux $\Phi$,
the density of states is a discrete sum of delta functions
at energies $E = (n + {1\over2}) \hbar \omega_c$, where
$n=0,1,2, ...$ is the Landau level index, and $\omega_c = eB/mc$ is
the cyclotron frequency.
Each Landau level has a degeneracy of $N = \Phi/\phi_0$, where
$\phi_0 = hc/e$ is the elementary flux quantum.

In the Landau gauge ${\bf A} = Bx{\hat {\bf y}}$, the single-particle
wave functions are given by
\begin{equation}
\varphi_{nk}(x,y) = {1 \over \sqrt{N}} e^{-iky}
H_n({{x-k{l_B}^2} \over {l_B}}) e^{-(x-kl_B^2)^2/2{l_B}^2} ,
\end{equation}
where $k=2\pi m/N$ indexes the states within a given Landau level,
$H_n$ denote Hermite polynomials, and the unit length is defined by
the magnetic length $l_B = \sqrt{\hbar c/eB}$.
The area $L^2$ of the system is measured by the number of
flux quanta, $N$, that pass through the sample, $L^2=2\pi l_B^2N$.
The Hamiltonian for the system in the LLL may be written as
\begin{equation}
H = \sum_{k,s} \sum_{k^\prime,s^\prime} |k,s\rangle
\langle k,s| (V+H_{SO}) |k^\prime,s^\prime\rangle
\langle k^\prime,s^\prime| ,
\end{equation}
where the basis set $\{ |k,s\rangle \}$ are the $n=0$
(LLL) eigenstates in Eq.~(3),
with spin quantum number $S_z = \hbar s$, where $s=\pm 1/2$.
The projected Hamiltonian in Eq.~(4) consists of both
a random Gaussian scalar potential $V$
and a spin-orbit scattering term $H_{SO}$.
The disorder correlation lengths of $V$ and $H_{SO}$
are taken to be the same, and are denoted by $\zeta$:
\begin{equation}
\overline{V({\bf r}) V({\bf r}^\prime)}
= U^2 {{e^{-|{\bf r}-{\bf r}^\prime|^2/{2\zeta^2}}}
\over {2 \pi \zeta^2}} ,
\end{equation}
where $U$ gives the strength of the scalar disorder.
The limit $\zeta\rightarrow0$
corresponds to uncorrelated white-noise disorder.
As we discuss in detail further below,
we take the SO scattering to have the form of a random
pseudofield ${\bf W}$ coupled to the electronic spin ${\bf \sigma}$:
\begin{equation}
H_{SO} = \lambda {\bf W} \cdot {\bbox\sigma} ,
\end{equation}
where $\lambda$ is a phenomenological SO coupling constant, and
\begin{equation}
\overline{W_\alpha({\bf r}) W_\beta({\bf r}^\prime)}
= \delta_{\alpha\beta} W^2
{{e^{-|{\bf r}-{\bf r}^\prime|^2/{2\zeta^2}}}
\over {2 \pi \zeta^2}} ,
\end{equation}
where $W$ gives the strength of the random pseudofield,
and $\alpha,\beta=x,y,z$.

Our numerical studies measure delocalization by
calculating the disorder-averaged Thouless number $T(E)$,
which has been argued to be proportional to the
disorder-averaged longitudinal conductivity
$\sigma_{xx}(E)$\cite{thou1}.
The proportionality of $T(E)$ and $\sigma_{xx}(E)$ has been
demonstrated, in the absence of a magnetic field, in Ref.\cite{akk}
In practice, we express the Hamiltonian as an $N \times N$ matrix,
which we diagonalize to obtain $N$ energy eigenvalues.
In Ref.\cite{thou1}, $T(E)$ is defined as the change $\Delta E$
in the energy eigenvalue at energy $E$ that results from
changing boundary conditions
({\it e.g.}, from periodic to antiperiodic),
multiplied by the total density of states $D(E)$.
The Thouless number is small for localized states because such states
do not extend to the boundaries.
It is relatively large for extended states because they
span the sample to its boundaries and hence are sensitive to changes
in the boundary conditions.
The width $\Delta E$ ({\it e.g.}, at half maximum) of $T(E)$
measures the energy range over which states are extended.
The width is finite for finite-size samples, but vanishes in
the limit that the sample size becomes infinite, since the extended
states for an infinite sample occur at discrete energies,
in the presence of a strong magnetic field.

The effect of changing the boundary conditions of wave functions is
equivalent to the gauge transformation that
results from adiabatically inserting flux
in a toroidal geometry\cite{haldane}.
This may be implemented {\it via} the perturbation Hamiltonian
\begin{equation}
\delta H = - {1 \over c} \int {\bf \delta A}
\cdot {\bf j} d{\bf r} ,
\end{equation}
where ${\bf \delta A}$ is the vector potential for a solenoid
carrying flux $\phi$.
Insertion of the flux $\phi=\phi_0/2$ is equivalent to changing the
boundary conditions from periodic to antiperiodic.
In the absence of a magnetic field,
the change in energy $\Delta E$ is second order in
$\delta H$\cite{thou1},
and $\Delta E$ may be calculated from second-order
perturbation theory in $\delta H$, which, like the Kubo formula for
the conductivity, has matrix elements involving the square of the
current operator.
Ando\cite{ando} has argued that the the derivation given
by Licciardello and Thouless\cite{thou1} relating the
Thouless number in the absence of a magnetic field can
also be applied to the case of a system in a strong magnetic field,
despite the lack of time reversal symmetry.

Reference\cite{thou1}
assumed that the mean-square value of the matrix elements
of the current operator between two states is not
too sensitive to the difference in energy between the states,
and that the energy levels are uncorrelated.
The latter assumption is not true
in the metallic regime\cite{altshu}.
Nonetheless, more recent work\cite{akk}
has extended the Thouless formula
to the metallic regime, and shown that the dissipative (transport)
conductance $g_d$ is proportional to the level curvature $g_c$,
defined as
\begin{equation}
g_c(E) = D(E) \Bigl\langle \left( {{\partial^2 E}\over
{\partial\phi^2}} \right)^2_{\phi=0} \Bigr\rangle^{1/2} .
\end{equation}
In the presence of a magnetic field, it is also
believed that $T(E)$ and $\sigma_{xx}(E)$ are proportional,
although we know of no proof of this.
Differences appear when a magnetic field is present;
for example, the energy shifts due to changing
the boundary conditions are linear rather than
quadratic in $\phi$, due to the lack of time reveral invariance,
and are random in sign.

The localization exponent $\nu$ is obtained by finite-size scaling.
We calculate the Thouless number for a range of samples sizes
$L \propto \sqrt{N}$.
In most of our calculations,
we have followed the approach of Ref.\cite{thou1}
and calculated the energy shifts between periodic
and antiperiodic boundary conditions,
studying $2\Delta E/\phi^2$ rather than
$\partial^2E/\partial\phi^2$, with $\phi=\phi_0/2$.
We expect that $T(E)$ should have the same scaling behavior
as $\sigma_{xx}(E)$.
{}For samples large enough to be in the scaling regime,
the Thouless number can be written
in terms of a scaling function of the ratio of
the localization length $\xi$ to the sample size $L$, $f(\xi/L)$.
Assuming that the localization length diverges at the LLL center
as $\xi \propto |E|^{-\nu}$,
we may write the Thouless number as a scaling
function of $|E| L^{1/\nu}$:
\begin{equation}
T(E) = f(\xi/L)
\sim f \lbrack (|E|L^{1/\nu})^{-\nu} \rbrack
\equiv \tilde{f}(|E|L^{1/\nu}) .
\end{equation}
The area $A(L)$ under the Thouless number curves
therefore scales like $L^{-1/\nu}$:
\begin{equation}
A(L) = \int_0^\infty T_L(E) dE
= L^{-1/\nu} \int_0^\infty \tilde{f}(E) dE
\equiv \tilde{f}_0 L^{-1/\nu} ,
\end{equation}
where $T(0)=\tilde{f}(0)$ and $\tilde{f}_0$
are independent of system size.
$T(E)$ should therefore have a half-width
$\Delta E \propto L^{-1/\nu}$
that narrows with increasing system size,
and a peak value $T(0)$, corresponding to the peak
conductivity $\sigma_{xx} = e^2/2h$\cite{peakc},
that is independent of the system size $L$.
We analyze our results by calculating $A(L)$
for each sample size $L$ and we check
that we are in the scaling regime by
ensuring that the Thouless number
data all scale onto a single curve
when the energy $E$ is rescaled to $E/A(L)$.
The scaling exponent $\nu$ can be obtained from
a log-log plot of $A(L)$ versus $L$, which, in the scaling regime,
should approach a straight line with slope $-1/\nu$.
Studying the scaling behavior of $A(L)$ has the advantage that
$A(L)$ is less noisy than $T(E)$,
since $A(L)$ is obtained by integrating $T(E)$.

{\bf 3. Spin-polarized electrons}

We first calculate the Thouless number for spin-polarized
electrons in the LLL, for white-noise disorder ($\zeta=0$).
The Thouless number plots for $N=20,80,300, 1000$ are shown in
{}Fig.~1a.
The half-width $\Delta E$ of the Thouless number
shrinks with increasing system size,
while the peak value remains constant at about $0.32 \approx 1/\pi$.
Ando's formula, $\sigma_{xx} = (e^2/4\hbar)T(E)$,
relating the Thouless number to the conductivity
for the case of a 2DEG in a strong magnetic field\cite{ando},
would imply that $T(0) \approx 1/\pi$
corresponds to a peak conductivity $\sigma_{xx}^c \approx e^2/2h$,
consistent with the known value for this model\cite{peakc}.
However, it should be noted that the value of the peak
Thouless number depends on the details of the method used to
calculate
the energy shift, and varies slightly with the correlation length
$\zeta$ of the disorder.
Nevertheless, we expect the peak value of $T(E)$ to be
independent of system size,
and the shape of the scaling function and the value of the
localization
exponent $\nu$ to be independent of the precise way the
Thouless number is defined.

We have checked that $T(E)$ is in the scaling regime by
rescaling the Thouless number plots for different system sizes
onto a single curve, shown in Fig.~1b,
for seven system sizes, from $N =$ 20 to 1000 flux quanta.
The data is collapsed onto a single curve by rescaling the energy
according to $E\rightarrow E/A(L)$.
A log-log plot of $A(L)$ versus $L$ is given in Fig.~1c.
A least-squares fit to a straight line gives a slope that corresponds
 to
$\nu = 2.36 \pm 0.02$ for the spin-polarized LLL
with white-noise disorder.
However, closer analysis shows that the data deviate from a
straight line, presumably due to finite-size corrections to
scaling\cite{huck},
and so we believe that the error estimate for our numerically
obtained
value of $\nu$ should significantly larger than $1\%$.
In any case, the value we obtain of $\nu\approx 2.4$
is consistent with previous numerical work
that used different methods and sample geometries
\cite{{huck},{huo},{liu},{chalk},{leedh},{mieck}}.

We have compared our numerical data for the Thouless number
to experimental longitudinal resistivity data,
taken from data used in Ref.\cite{wei2}.
{}Figure 2a shows the experimental longitudinal resistivity data
versus applied magnetic field,
for an InGaAs/InP heterostructure at four temperatures,
T = 40, 110, 305, 640mK.
The symbols in Fig.~2b are the data of Fig.~2a,
centered, normalized by the peak resistivity, and rescaled,
for four different temperatures\cite{wei4}.
The width of $\rho_{xx}(B)$ narrows as the temperature
becomes smaller.
It does so as a power law of the temperature, with associated
exponent $\kappa\approx0.42\pm0.04$.
Under rescaling of the magnetic field with
the temperature according to
$\Delta B \rightarrow \Delta B \cdot T^{-\kappa}$,
all the longitudinal resistivity curves can be collapsed onto
a single curve.

{}Figure~2b shows that the scaled Thouless number data
fits the scaled resistivity data.
The fit between the temperature-rescaled $\rho_{xx}$ data
and the size-rescaled $T(E)$ curves is further evidence
that the temperature and length are related by an
inelastic scattering exponent value of $p=2$,
which has also been found in other experiments\cite{wei3}.
{}Figure~2b is the universal crossover function
describing the transition from metal to insulator.
Reference\cite{leedh} also used the network model to
calculate a crossover function
as a function of the localization length $\xi$,
and obtained good agreement with the experiments of
Ref.\cite{mceuen},
after correcting for the contribution due to edge currents.
Experiments of Hughes {\it et al.}\cite{hughes}
find single-parameter scaling of $\sigma_{xx}(B,T)$,
with $\kappa \approx 0.42$ and
$\sigma_{xx}$ well fit by the model of Ref.\cite{leedh}.
The experimental data in Fig.~2 were obtained from samples of width
600 microns; we expect edge currents in these samples to be less
important
than for the samples used in Ref.\cite{mceuen}, which were only 7
microns wide.
If Coulomb interactions dominate,
one estimate for the sample size at which edge effects become
important
is when the sample width is smaller than the electron scattering
length
$l_T$, where $e^2/\epsilon l_T \sim k_BT$.
{}For T=20mK, this gives $l_T \sim 50\mu m$.

{\bf 4. Spin-unresolved quantum Hall transition}

Reference~\cite{hwang} studied the scaling of the maximum slope of
the
Hall resistivity as a function of temperature, for spin-unresolved
quantum Hall transitions.
Recall that for the spin-resolved Landau levels the scaling exponent
is
$\kappa=0.42$\cite{wei1};
when the spin-splitting is not resolved, it is found that
$\kappa=0.21$\cite{{wei1},{hwang}}.
This would correspond to doubling the exponent $\nu \approx 2.3$
to $\nu \approx 4.6$,
assuming that the inelastic scattering length exponent $p$ is
unchanged.
Other experiments\cite{wei3} give evidence that $p=2$ for both the
spin-resolved and spin-unresolved cases.

In practice, electrons can be spin-flip scattered,
even in the absence of
magnetic impurities, due to the spin-orbit (SO) interaction.
The SO interaction is a relativistic effect, and is therefore
small in comparison to the random potential scattering.
Nonetheless such scattering could be important
when the spin levels are degenerate,
and is stronger for materials with scatterers of high atomic number,
such as the InGaAs samples used in Ref.\cite{hwang}.

SO scattering is due to the fact that an electron moving
with velocity ${\bf v}$ in the presence
of an electric field ${\bf E} = {1 \over e} {\bbox\nabla} V$
experiences in its rest frame an effective magnetic field
${\bf B} = {1 \over c} {\bf v} {\bbox\times} {\bf E}$
which couples to the electron spin according to
$(e\hbar/mc) {\bf B} \cdot ({\hbar \over 2}) {\bbox\sigma}$.
The orbital motion of an electron therefore gives rise to
an effective magnetic field that couples to the spin.
The SO interaction is written as
\begin{equation}
V_{\rm SO} = {\hbar\over{4m^2c^2}} {\bbox\sigma}
\cdot {\bbox\nabla}V {\bbox\times\Pi} ,
\end{equation}
where ${\bf \Pi} = -i \hbar {\bbox\nabla} + (e/c){\bf A}$
is the mechanical momentum of the electron,
whose spin is altered by the effect of the Pauli matrices
${\bf \sigma}$ in Eq.~(13).

The effect of $V_{\rm SO}$ on the Bloch electrons of
bulk III-V compounds such as GaAs
has been investigated in Refs.\cite{A3B5,pikus}.
In the vicinity of the $\Gamma$
point, the effective conduction band Hamiltonian becomes\cite{pikus}
\begin{equation}
H_{\rm eff}={\hbar^2{\bf k}^2\over 2 m^*} +
{\widetilde\beta} {\bbox\sigma}\cdot{\bbox\kappa}
\end{equation}
where $m^*$ is the conduction band effective mass,
$\kappa_x=k_x^{\vphantom{2}}(k_z^2-k_y^2)$,
and $\kappa_y$ and $\kappa_z$ are obtained from the expression for
$\kappa_x$ by cyclic permutation.
In a material that lacks inversion symmetry,
the SO interaction [Eq.~(13)] leads to a nonzero value of the
constant
${\widetilde\beta}$, and hence to a splitting of the conduction
band states which is proportional
to the cube of the wave vector.
Application of Eq.~(14) to inversion layers, developed in
Refs.\cite{khae,romanov}, leads to the following effective SO
Hamiltonian:
\begin{equation}
H_{\rm SO}={{\bf p}^2\over 2 m^*}
+\alpha(\sigma^x p_y - \sigma^y p_x)
+\beta (\sigma^x p_x - \sigma^y p_y)
\end{equation}
The third term is identified as
${\widetilde\beta}\langle{\bbox\sigma}\cdot{\bbox\kappa} \rangle$,
the expectation value of the effective conduction-band SO interaction
taken in the appropriate quantum-well bound state
(the interface is assumed normal to ${\hat{\bf z}}$).
Khaetskii\cite{khae} finds
$\beta\simeq 1.1\times 10^{5}$ cm/s
for an inversion layer density of $n=2\times 10^{11}$ cm$^{-2}$.
The second term arises due to the asymmetry of the confining
potential,
which generates an electric field in the inversion layer.
Only rough estimates of the constant $\alpha$ exist\cite{khae,bych},
with $\alpha$ assumed to be in the range $0.1\beta$ to $\beta$.
In any case,
for typical momenta of size $\hbar/l_B$, where $l_B\sim 100\AA$,
the SO energy scale is roughly $10^{-5}$ eV, or 0.1 - 1.0 K.

The SO Hamiltonian in Eq.~(15) produces random scattering of the
spin in the presence of ordinary random scalar potential scattering.
This is because scalar scattering changes the momentum of the
scattered electron, which by Eq.~(15) is coupled to its spin.
We estimate direct SO scattering from
the slowly varying impurity potentials
in high mobility GaAs heterojunctions to be extremely small
and unlikely to have any significant effect\cite{footsmg}.
However, in the InGaAs samples studied in Ref.\cite{wei2}
(which have a mobility of 35,000 cm$^2$/Vs),
alloy scattering from short-range potential disorder
produces larger momentum transfers and hence larger spin-flip
amplitudes.
In addition, the relatively large atomic number of In may further
enhance SO scattering.
Nevertheless, the SO energy scale is much smaller than that
for scalar potential disorder.

In the presence of an external magnetic field
${\bf B}={\bbox\nabla\times}{\bf A}=B{\hat{\bf z}}$,
one must add a Zeeman term
$H_{\rm Z}=-\frac{1}{2} g \mu_{\rm B} B\sigma^z$, as well as make the
substitution ${\bf p}\to
{\bbox\Pi}\equiv -i\hbar{\bbox\nabla} +(e/c){\bf A}$.
We have included electron spin in our Thouless number calculations,
in the limit of zero Zeeman splitting ($g=0$),
so that the spins are not resolved.
Since ${\bf \Pi}$ is not a constant of motion for
an electron moving along an equipotential contour of
the random potential $V$,
we take, for simplicity, the SO coupling to have the form of Eq.~(7),
a Zeeman coupling to a random pseudofield ${\bf W}$.
In the absence of spin-orbit scattering,
the exponent $\nu$ remains the same as for the spin-resolved case,
although $T(E)$ doubles,
reflecting the doubling of the density of states due to the spin
degeneracy.

The use of Eq.~(7) as the SO interaction rather than the more
cumbersome Eq.~(15) can be motivated by considering
the Hamiltonian $H = H_0 + V$, where V is a random scalar potential,
and
\begin{equation}
H_0 = {{{\bf \Pi}^2}\over{2m^*}} + \beta
{\bbox\Pi}\cdot{\bbox\sigma}.
\end{equation}
This is just Eq.~(15) in the presence of a magnetic field,
and in the absence of the term proportional to $\alpha$.
(For convenience, we have performed a $\pi$ rotation about the
$x$-axis
in spin space to change $-\sigma_y \rightarrow \sigma_y$ in
Eq.~(15).)
$H_0$ may be written as
\begin{equation}
H_0 = \hbar\omega_c (a^\dagger a + 1/2)
+ {{\hbar\beta}\over{\sqrt{2}l_B}} (a^\dagger\sigma_- + a\sigma_+) ,
\end{equation}
where the lowering operator is defined as
\begin{equation}
a = {{l_B}\over{\sqrt{2}\hbar}} (\Pi_x - i\Pi_y) ,
\end{equation}
the raising operator is the hermitian conjugate of $a$,
and $\sigma_\pm = \sigma_x \pm i\sigma_y$.
$H_0$ can be diagonalized exactly, and has eigenstates of the form
\begin{eqnarray}
|k+\rangle &=& u_+ |2k,\uparrow\rangle + v_+
|2k+1,\downarrow\rangle\\
\nonumber
|k-\rangle &=& u_- |2k,\downarrow\rangle
+ v_- |2k-1,\uparrow\rangle ,
\end{eqnarray}
where $|n,s\rangle$ denotes a state with Landau level index $n$ and
spin $s$ ($s = \uparrow , \downarrow)$.
We have omitted the indices for the $N$-fold degenerate states within
each Landau level.

We now restrict our consideration to the energy eigenstates
of $H_0$ with the two lowest energies.
The other eigenstates are higher in energy by
multiples of $\hbar\omega_c$.
The lowest energy states are
\begin{eqnarray}
\matrix{|+\rangle = |0,\downarrow\rangle,&E_+=0\cr
	|-\rangle = (1-\gamma^2/2)|0,\uparrow\rangle
		    - \gamma |1,\downarrow\rangle,
		    &E_-=-\gamma^2\hbar\omega_c\cr} ,
\end{eqnarray}
where $\gamma\equiv\beta/\sqrt{2}l_B\omega_c\sim10^{-4}$,
and we measure the energies $E_\pm$ with respect to
$\hbar\omega_c/2$.

In the subspace of these states,
it follows from Eq.~(19) that the effective scattering Hamiltonian
can be rewritten, up to a constant term, as
\begin{equation}
H_{eff} = \widetilde{V} + \lambda{\bf W}\cdot{\bbox\sigma} ,
\end{equation}
where $\widetilde{V}$ is just the scalar potential $V$ projected
onto the LLL, and, to order $\gamma$,
\begin{equation}
(W_x,W_y) = -2\gamma
(\rm{Re}\langle +|V|-\rangle,
\rm{Im}\langle +|V|-\rangle) .
\end{equation}
$W_z$ is of order $\gamma^2$, but is not expected to
change the universality class of the delocalization transition
in the presence of the scalar potential $V$ which, like a
$W_z\sigma_z$ term,
produces scattering without flipping the spin.

If SO scattering were a relevant perturbation (one that changes the
universality class of the transition), its effects would become large
at sufficiently long length scales and low enough temperatures.
We check the relevance of SO scattering by
artificially increasing the strength of the SO scattering,
and then calculating the exponent $\nu$ and the peak value of the
Thouless number $T(0)$.
When short-range SO scattering is included, we find that the
half-width
$\Delta E$ of the Thouless number curve becomes larger
and has a scaling exponent
of about $\nu=4.4\pm 0.2$, assuming that the localization length
diverges
only at the band center, as for the spin-resolved case.
This is illustrated in Fig.~3a by the Thouless number curves for
system sizes
of $N$ = 20, 80, and 500, which are peaked at $E=0$.
{}Figure~3b shows a log-log plot of $A(L)$ versus $L$,
as in Fig.~1c.
The apparent value of $\nu$ is twice that of the spin-resolved value,
in agreement with the experimental results of Ref.\cite{hwang}.
The peak value of the Thouless number, $T(0)$, increases from
0.32 to roughly 0.4.
This is similar to the behavior seen in the experiments of
Ref.~\cite{hwang}
and in the numerical simulations of Ref.\cite{chalk},
when $\xi$ is assumed to diverge at a single energy.
This would seem to suggest that SO scattering might belong to
a different universality class than ordinary scalar potential
scattering.

However, when the SO scattering potential is smoothed ($\zeta=2$),
we obtain very different results.
This is illustrated in Fig.~4a, where we have plotted the Thouless
number
for system sizes of $N$ = 40, 160, and 500.
The Thouless number curves acquire two symmetrically displaced peaks
at
$\pm E_c$ (we show only one, for $E_c>0$),
each with a peak value of approximately $1/\pi$, just as for the
spin-resolved case in the absence of SO scattering.
If we assume that the Thouless number is the sum of symmetrically
displaced
scaling functions,
then a log-log plot of $A(L)$ versus $L$,
shown in Figure~4b, yields $\nu\approx 2.5$,
much closer to the value for the spin-resolved exponent, $\nu=2.3$.
The similarity of the peak height $T(0)$ and the exponent $\nu$
to the case of spinless electrons are evidence
that the universality class of the
delocalization transition is the same for the spin-unresolved
and spinless non-interacting electrons.

We have also varied the strength $\lambda$ of the SO scattering
to measure its effect on the energy $E_c$ at which $T(E)$ peaks.
This is shown in Fig.~5a, where we have plotted $T(E)$ versus $E$ for
$N=80$ and $\zeta=2$, with $\lambda=0,2,4,6$.
We find that for this simple form of SO scattering,
$E_c \propto \langle \lambda |{\bf W}| \rangle$,
just as for Zeeman splitting.
{}Figure~5a can be understood in terms of an effective Zeeman coupling
generated
by the smooth SO scattering\cite{leedh}.
Denoting the correlation length of the SO scattering by $\zeta$,
and the average strength of the random field by $\langle |{\bf W}|
\rangle$,
it can be shown that when $\zeta \gg l_B$,
so that ${\bf W}$ is slowly varying,
the SO field ${\bf W}$ acts like a local constant magnetic field
coupling to the spins and generates an effective Zeeman splitting
that depends on the magnitude (but not the direction) of ${\bf W}$.
The slow variation in the direction of ${\bf W}$ yields a Berry's
phase
that acts like a weak random magnetic field, but this does
not change the universality class in the presence of a strong
external magnetic field\cite{wang,leedh2}.
The net effect, therefore, is to produce a splitting between
$\pm E_c$ of magnitude $\Delta E = 2 \langle \lambda |{\bf W}|
\rangle$.
This behavior is shown in Figs.~5b and 6b.

It has been suggested that in some IQHE models, the universality
class of the
metal-insulator transition for white-noise disorder ($\zeta=0$) is
different
than for smooth disorder\cite{liu}.
In order to investigate whether this is the case here,
we have artificially increased the value of $\lambda$
as in Fig.~5a, except that we use white-noise disorder.
The results for $N=20$, with $\lambda=0,2,4,6$ are shown in Fig.~6a.
It can be seen that, as in the case of smooth disorder (Fig.~5a), the
 peak of
the Thouless number is displaced away from zero ($E_c\not=0$).
A log-log plot of $A(L)$ versus $L$, however, yields $\nu=4.4\pm
0.2$.
We believe this is due to the slow crossover to the scaling regime
when $\zeta/l_B<1$, as found by Huckestein\cite{huck},
but cannot demonstrate this within the range of available sample
sizes.

Huckestein has studied numerically finite-size corrections to scaling
in the center of Landau levels\cite{huck}.
Although the value of $\nu\approx 2.3$ is obtained
for the lowest $(n=0)$ LL, independent of the
correlation length $\zeta$ of the disorder potential,
this is not the case for the second ($n=1$) LL.
{}For $n=1$, $\nu\approx 2.3$ is observed only when $\zeta\ge l_B$.
{}For $\zeta<l_B$ no universal scaling behavior was observed.
{}Following Chalker and Eastmond\cite{chalkeast}, Huckestein
showed that deviations from finite-size scaling themselves scaled
with size, so that, in effect, Eq.~(10) is modified to
\begin{equation}
T(E) \sim \tilde{f}(|E|L^{1/\nu},l_{irr} L^{-y_{irr}}) ,
\end{equation}
where $l_{irr}$ is a length scale that is a function of $\zeta/l_B$.
The exponent $y_{irr}\approx0.38\pm0.04$
obtained by Huckestein\cite{huck} agrees with the value obtained
in Ref.\cite{chalkeast} for network model simulations.
Of great practical importance is the fact that the size of the
length scale $l_{irr}$ is $10^4$ times longer for $\zeta=0$
than for $\zeta=0.8l_B$.
This makes it difficult to observe scaling behavior for $n=1$
when $\zeta=0$.
It is possible that the apparent dependence of
$\nu$ on $\zeta$ that we find in the spin-degenerate case may
have a similar origin to the difficulties encountered
in reaching the scaling regime when $n=1$.

{\bf 5. HSW model}

Hikami, Shirai, and Wegner (HSW) have considered a special model for
Anderson localization of spin-degenerate non-interacting
electrons\cite{hsw}.
They study the case when only scattering between one spin state and
a {\it different} spin state is allowed.
This corresponds to taking the pseudofield ${\bf W}$ to lie in the
$xy$
plane, and completely supressing ordinary scalar potential
scattering.
Although it excludes important scattering processes present in real
samples,
it is the simplest model of spin-orbit scattering in the quantum Hall
 regime,
and has the great advantage of being analytically tractable
for the case of a Gaussian white-noise random distribution of
scatterers\cite{hsw}.

The model of HSW exhibits novel behavior near the band center
($E=0$).
The conductivity at $E=0$ is given exactly by $e^2/\pi h$,
and the DOS diverges at the band center\cite{hsw,gade}.
HSW used a $1/S$ expansion,
where $S$ is essentially the electron spin (physically, $S=1/2$),
to calculate the DOS to order $1/S^2$.
They found a $1/S^2$ enhancement of the DOS at $E=0$ of the form
$(1/S^2) \ln^2(|E/V_{\rm{RMS}}|)$.
Using an effective field theory describing the IQHE and
projecting on to the LLL,
Gade\cite{gade} finds that the DOS diverges instead as a power law,
$D(E) \propto |E|^{-\delta}$, near $E=0$.
We also note that Ludwig {\it et al.}\cite{ludwig}
have studied a class of IQHE models
which respect symmetries analogous to those of the HSW model,
and that they also obtain a DOS that diverges as a power law near
$E=0$.

We have calculated the DOS and Thouless number for the HSW model,
and we find that both quantities vanish at $E=0$, for
finite-size samples.
{}Figure~7a shows the DOS of the HSW model as a function of
energy for the small system size of $N$=10 flux quanta.
System sizes of $N$=20,40,80,160 also show the same behavior;
namely, a DOS with two strong peaks near $E=E_0$ due to
the energy level spacing in a finite sample, where
$E_0 \propto V_{\rm{RMS}}/N$,
corresponding to the average level spacing between energy
eigenvalues.
In addition, the DOS vanishes linearly as $E\rightarrow0$, for
$E<E_0$.
We find that the peak DOS at $E=E_0$ increases with $N$.
The HSW Hamiltonian may be written in the form
\begin{equation}
H_{HSW} = \widetilde{W_x}\sigma_x + \widetilde{W_y}\sigma_y ,
\end{equation}
where $\widetilde{W_\alpha}$ denotes the components of the
pseudofield
${\bf W}$, projected onto the LLL,
and the $\sigma_\alpha$ are Pauli matrices.
Note that $\sigma_z (H_{HSW}) \sigma_z = -H_{HSW}$,
so that for every eigenstate $|\psi\rangle$ of energy $E$,
$\sigma_z|\psi\rangle$ is also an eigenvalue, with energy
$-E$\cite{leedkk2}.
The energy spectrum of $H_{HSW}$ therefore possesses a reflection
symmetry
about $E=0$.
The reduction of the DOS near $E=0$ is due to the combined effects of
the reflection symmetry of the eigenvalue spectrum,
plus the usual level repulsion of eigenvalues.
It is therefore a finite-size effect.
We note that when the disorder is smoothed ($\zeta>0$),
the DOS becomes more strongly peaked near $E=0$,
and that the downturn moves closer to $E=0$ ($E_0$ becomes smaller).
This is illustrated in Fig.~7b, where we have plotted the DOS
for a samples of size $N=80$, for disorder correlation lengths
$\zeta$=0,1,2 (in units of the magnetic length).
The case of smooth disorder was studied in Ref.\cite{leedkk2}.

In order to observe a divergence in the DOS near $E=0$, one must
study
energies small enough that the HSW correction can be observed,
but large enough ($E>V_{\rm{RMS}}/N$) that the vanishing of the DOS
due to eigenvalue repulsion does not interfere.
Given our numerical limitations on the maximum size ($N_{max}\approx
500$)
of our samples, we were unable to isolate any divergent contribution
to the DOS at small $E$.
To better understand the behavior of the HSW model near $E=0$,
we studied a random matrix problem with the same symmetry
as the HSW Hamiltonian, namely the eigenvalue spectrum for
matrices with the same form as the HSW Hamiltonian,
\begin{eqnarray}
H=\left(\matrix{0&W\cr
		W^\dagger&0\cr}\right) ,
\end{eqnarray}
where $W$ is a Gaussian random complex matrix\cite{mehta}.
When the elements $W_{ij}$ all have the same variance, so that
\begin{equation}
P(W_{ij}) \propto e^{-|W_{ij}|^2/2\sigma^2} ,
\end{equation}
one obtains the exponential ensemble of random matrices,
studied by Bronk\cite{bronk}.
The joint probability distribution function
for the eigenvalues is\cite{bronk}
\begin{equation}
P(E_1,...,E_N) \propto \prod_{i} |E_i| \prod_{j<k} |E_j^2 - E_k^2|^2
e^{-\sum_i E_i^2/2\sigma^2} .
\end{equation}
Using a mathematical identity for sums of the product
of two Laguerre polynomials\cite{gr}, one can show that the DOS for
the exponential ensemble is given by
\begin{eqnarray}
D(E_1) = \int dE_2 ... \int dE_N P(E_1,...,E_N)
\nonumber \\
\propto |E_1|
\left[ L_{N-1}^{(1)}(x) L_{N-1}^{(0)}(x)
- L_{N-2}^{(1)}(x) L_{N}^{(0)}(x) \right]
e^{-x} ,
\end{eqnarray}
where $x=E_1^2/2\sigma^2$ and the $L_n^{(\alpha)}$
denote associated Laguerre polynomials.
The DOS is proportional to $|E_1|$ near $E_1=0$,
and vanishes like $e^{-E_1^2/2\sigma^2}$ for large $E_1$,
which is the behavior seen in Fig.~7a for the HSW DOS.
However, Bronk's\cite{bronk} calculation of
the asymptotic ($N\rightarrow\infty$)
random-matrix DOS may be invoked
to show that the usual semicircle law for the DOS holds,
so that the random matrix model gives
no divergent contributions to the DOS.
Not surprisingly, the HSW model and the exponential ensemble have
very different behavior even near $E=0$, despite the similar
symmetries of the Hamiltonians.
Random matrix theory is expected to apply in
metallic phases but not at critical points\cite{metalrm}.

{}Figure~7c shows the Thouless number for the HSW model as a function
of energy for system sizes $N$=10,40,160.
It can be seen from Fig.~7c that there are two delocalization
transitions, one at $E_c\sim 0.7V_{\rm{RMS}}$, and the other at
$E=0$.
The transition at $E_c\sim 0.7V_{\rm{RMS}}$ is reminiscent of the
transition
at $E\sim\lambda|{\bf W}|$ studied in Sec.~4,
for strong SO scattering.
As the system size increases, it is clearly seen that
the peak in $T(E)$ at $E=E_c\sim 0.7V_{\rm{RMS}}$ becomes narrower
and
more pronounced.
We also find that the peak value, $T(E_c)$, for the HSW model
is close (only 10\% smaller) to that for spinless electrons, $T(0)$.
The transition at $E=0$ is more difficult to characterize,
since the width of the peak in $T(E)$ near $E=0$ seems to scale
like the energy eigenvalue spacing, {\it i.e.},
$\Delta E \propto 1/N \propto L^{-2}$,
rather than $\Delta E \propto L^{-3/7}$ (near $E_c$).
The delocalization transition near $E=0$ also differs from
the one at $E_c$ in that
the peak Thouless number is seen to (slowly) decrease
with increasing system size.
These differences in the behavior of $T(E)$ are evidence that the
universality class of the two delocalization transitions
in the HSW model are different.

{\bf 6. Conclusions}

Using a continuum LLL model, we find,
in agreement with previous numerical work, that for spin-polarized
electrons in the LLL, the localization exponent has the value
$\nu\approx2.4$, and that the peak value of the Thouless number
is constant, independent of the sample size.
We find that the scaled Thouless-number data fits the scaled
experimental
resistivity data of Ref.\cite{wei4}, which is the universal crossover
function for the quantum Hall metal-insulator transition.

{}For short-range SO scattering, we find, as in the experiments of
Ref.\cite{hwang}, a seemingly doubled exponent, $\nu \approx 4.4$,
when we assume that the Thouless number is peaked at a single energy.
This value of $\nu$ reverts to close to the usual non-SO value of
$\nu \approx 2.3$, and the peak value of the Thouless number
approximates the spin-polarized value,
when the SO scattering potential is smoothed
and made sufficiently large.
We conclude that the universality class of the metal-insulator
transition
for non-interacting spin-degenerate electrons in the quantum Hall
regime
is the same as for spin-polarized electrons.

It is possible that the experiments leading to an apparently
different value
of the temperature exponent $\kappa$ may not be in the scaling
regime,
especially for nearly white-noise scattering, which is the situation
for
the In-doped GaAs samples used in Ref.\cite{hwang}.
The experiments of Ref.\cite{hwang} required temperatures $T>50$mK;
otherwise, the spin-splitting between the Landau levels was resolved,
and the putative spin-degeneracy was lost.
It could be that the effective sample size
$l_{in} \propto T^{-p}$ was too short to reach the scaling regime,
even at the low temperatures used in the experiments.

It is not known whether Coulomb interactions change the value of
$\kappa$
for spin-degenerate electrons, {\it e.g.}, by changing the
inelastic scattering length exponent $p$ (or the dynamical exponent
$z$)
from its value for non-interacting electrons\cite{sondhi}.
It is believed that for dirty boson systems, the dynamical exponent
depends on the form of the interparticle potential\cite{dirtyb}.
Certainly a peak value of the longitudinal conductivity less than
the universal value $\sigma_{xx}^c=e^2/2h$ for non-interacting
electrons
is seen in experiments, and perhaps this is due to Coulomb
interactions.
The importance of Coulomb interactions even for the spin-polarized
case is underscored by the
large exchange enhancement of the elecron $g$ factor
relative to its bare value, which has been studied in numerous
experiments\cite{gexp} and calculations\cite{gthe}.
Recent work has developed a new picture of the
nature of spin excitations in the quantum Hall regime,
in the presence of Coulomb interactions\cite{skyrmion}.
In principle, the $1/r$ nature of the Coulomb interaction in a 2DEG
could be purposely modified by specially gating a heterojunction
with a metallic overlayer, so as to screen the Coulomb interaction
and induce $1/r^3$ dipole-dipole interparticle interactions.
This has been carried out for the insulating regime, where hopping
dominates the transport\cite{hopthe,hopexp}.

We also studied the model of Hikami, Shirai, and Wegner\cite{hsw},
and calculated the DOS and Thouless number for a range of system
sizes.
We found that the DOS and $T(E)$ both vanish linearly at $E=0$ for
finite-size
samples, contrary to the naive expectation that the DOS should
increase
and $T(E)$ become constant at $E=0$.
The different scaling behavior of the Thouless number at the two
energies
are evidence that the two delocalization transitions are in different
universality classes.


\acknowledgments

This research was  supported by the National Science Foundation
through grant NSF DMR-9416906 at Indiana University,
and by NSF grant 89-57993 at UC San Diego.
We are grateful to C.~Canali, J.~Chalker, S.~Das~Sarma, M.~Guo,
R.~Hyman, A.~H.~MacDonald, and S.~Sondhi for helpful discussions,
and to H.~P.~Wei and D.~C.~Tsui for the data used in Fig.~2.
\\

\begin{figure}

\caption[]{ (a) Thouless number data for spin-polarized electrons
in the lowest Landau level, in the presence of white-noise disorder
($\zeta=0$), for sample sizes of $N$=20, 80, 300, 1000 flux quanta.
The width of the Thouless number decreases with increasing $N$.
\hfill \break
(b) The scaled Thouless number data
for spin-polarized electrons in the lowest Landau level,
for $N$=20, 40, 80, 160, 300, 600, 1000 flux quanta,
falls onto a single curve.
\hfill \break
(c) Log-log plot of the integrated area under the Thouless number
curves, versus the system size ($L \propto \sqrt{N}$),
for $N$=20, 40, 80, 160, 300, 600, 1000 flux quanta.
The slope of the resulting line is $-1/\nu$, and yields
$\nu \approx 2.4$.
\label{jumppic1}}
\hfill \break

\caption[]{ (a) Longitudinal resistivity ($\rho_{xx}$) data versus
applied magnetic field for four temperatures,
$T$=40, 110, 305, 640 mK,
taken from Ref.\cite{wei2}.
\hfill \break
(b) Scaled, normalized experimental data for $\rho_{xx}$
for $T=40, 110, 305, 640$ mK, shown by symbols.
The solid lines are the scaled Thouless number data
for spin-polarized electrons in the lowest Landau level,
for $N$=20, 40, 80, 160, 300, 600, 1000 flux quanta.
This curve represents the universal crossover function from metal
to insulator for non-interacting electrons in the quantum Hall
regime.
\label{jumppic2}}
\hfill \break

\caption[]{ (a) Thouless number data with spin-orbit scattering,
white-noise disorder ($\zeta=0$), and sample sizes of
$N$=20,80,500 flux quanta.
The width of the Thouless number decreases with increasing $N$.
\hfill \break
(b) Log-log plot of the integrated area under the Thouless number
curves, versus the system size.
The slope of the resulting line yields $\nu \approx 4.4$.
\label{jumppic5}}
\hfill \break

\caption[]{ (a) Thouless number data for smooth spin-orbit
scattering,
for smooth ($\zeta=2$) disorder, and sample sizes of
$N$=40, 160, 500 flux quanta.
Note that the data are peaked away from zero energy, at the same
energy $E_c$.
The width of the Thouless number decreases with increasing $N$.
\hfill \break
(b) Log-log plot of the integrated area under the Thouless number
curves, versus the system size.
The slope of the resulting line yields $\nu \approx 2.5$.
\label{jumppic6}}
\hfill \break

\caption[]{ (a) Thouless number data for smooth spin-orbit scattering
($\zeta=2$), for spin-orbit coupling values of $\lambda$=0, 2, 4, 6.
The Thouless number is peaked at
$E_c \propto \langle \lambda |{\bf W}| \rangle$.
\hfill \break
(b) $E_c$ versus $\lambda$, for $\lambda$=0, 1, 2, 3, 4, 5, 6.
\label{jumppic7}}
\hfill \break

\caption[]{ (a) Thouless number data for strong spin-orbit scattering
($\lambda \langle | {\bf W} | \rangle \sim V_{\rm{RMS}} $)
and white-noise disorder ($\zeta=0$),
for spin-orbit coupling values of $\lambda$=0, 2, 4, 6.
\hfill \break
(b) $E_c$ versus $\lambda$, for $\lambda$=0, 1, 2, 3, 4, 5, 6.
\label{jumppic8}}
\hfill \break

\caption[]{ (a) Density of states $D(E)$ for pure spin-orbit
scattering model of Hikame, Shirai, and Wegner (HSW)\cite{hsw},
for system size $N$=10 flux quanta.
{}For finite $N$, $D(E=0)=0$.
$D(E)$ exhibits a strong peak at an energy $E_0 \propto 1/N$,
close to $E=0$,
with a peak height that grows slowly with $N$.
\hfill \break
(b) Density of states for HSW model, for fixed $N$=80, and varying
disorder correlation lengths $\zeta=0,1,2$.
The peak DOS near $E=0$ increases with $\zeta$.
\hfill \break
(c) Thouless number $T(E)$ for HSW model, for system sizes
$N=10, 40, 160$,
showing two delocalization transitions, at $E\sim 0.7V_{\rm{RMS}}$
and
$E=0$.
{}For finite $N$, $T(E=0)=0$.
The Thouless number decreases with increasing $N$, except near
the fixed points $E\approx\pm 0.7V_{\rm{RMS}}$.
\label{jumppic9}}
\hfill \break

\end{figure}


\begin{references}
\bibitem{footsmg}
Spin-orbit scattering from the short-range alloy potential
of the heavy (high-$Z$) In in the InGaAs samples
may be more significant than in high mobility GaAs heterojunctions,
but its magnitude is difficult to estimate accurately.
\end{references}
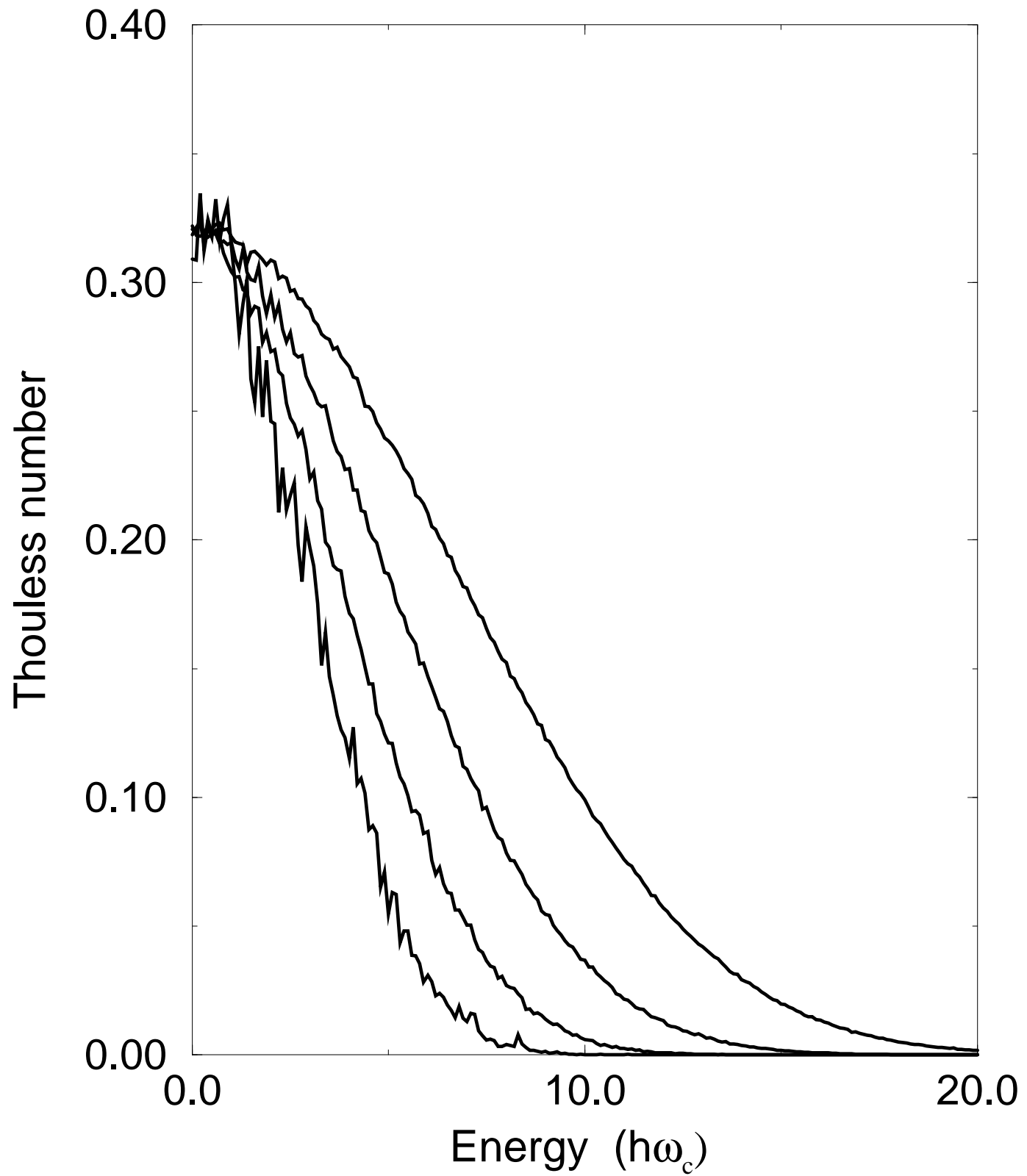
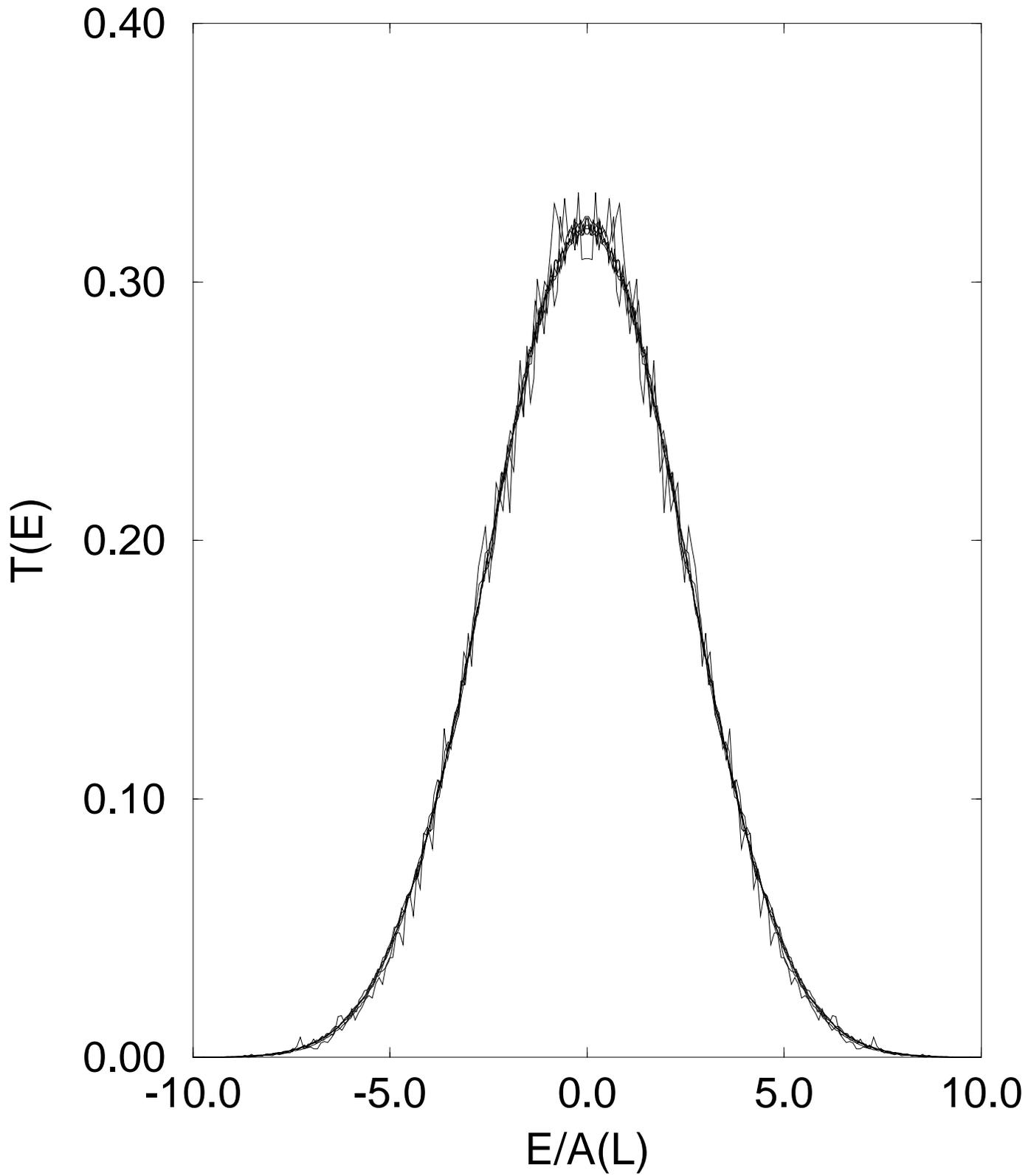
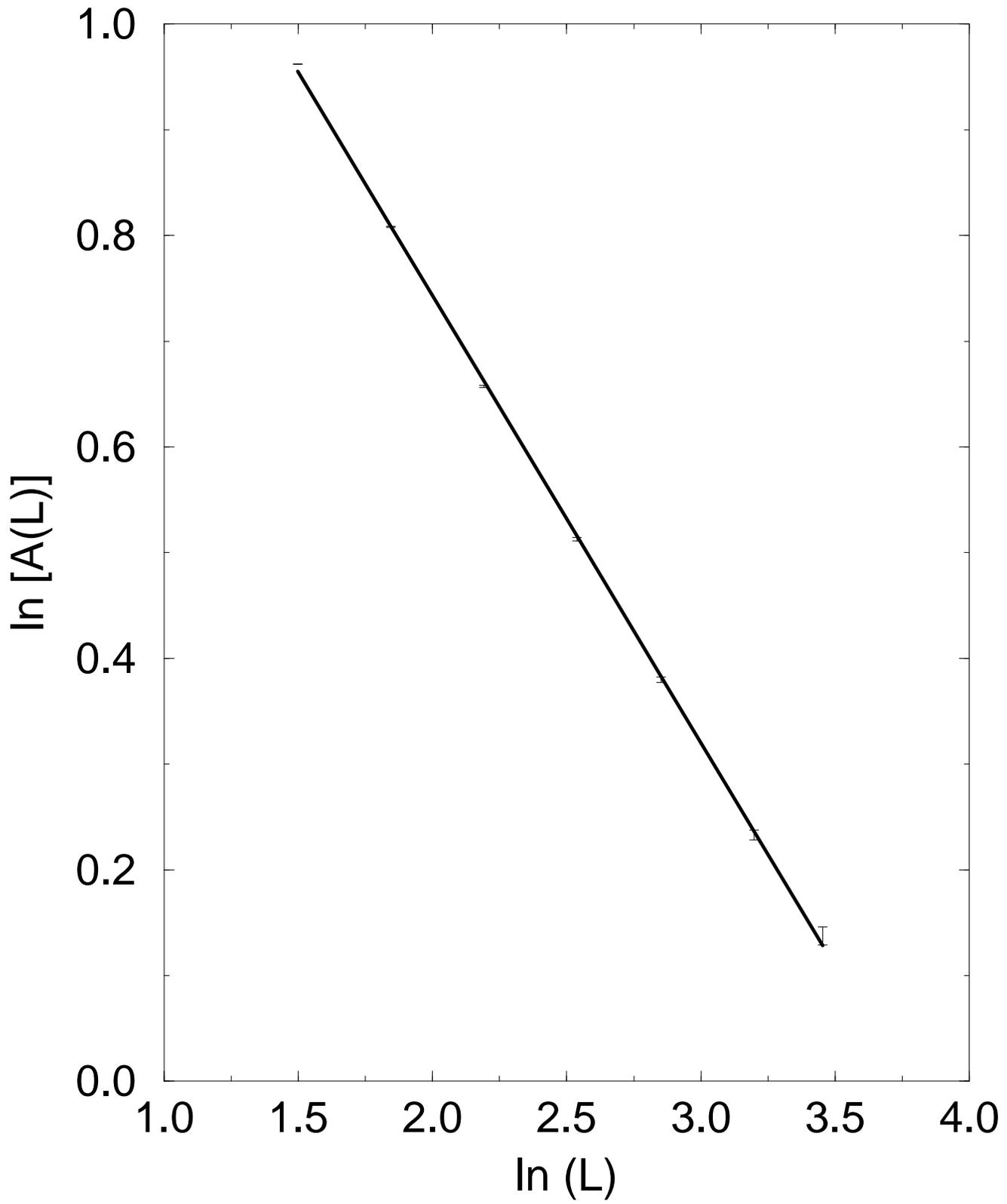
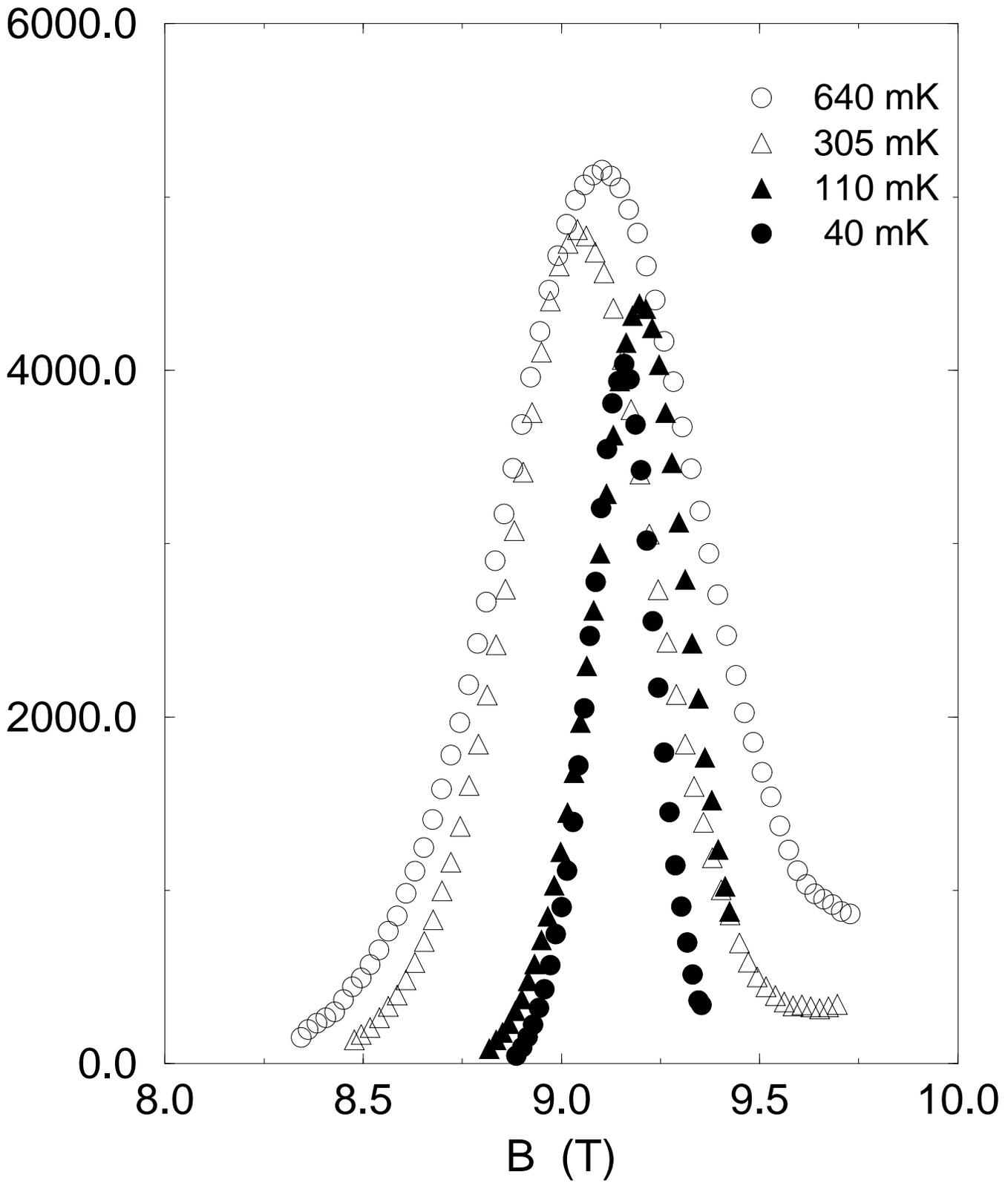
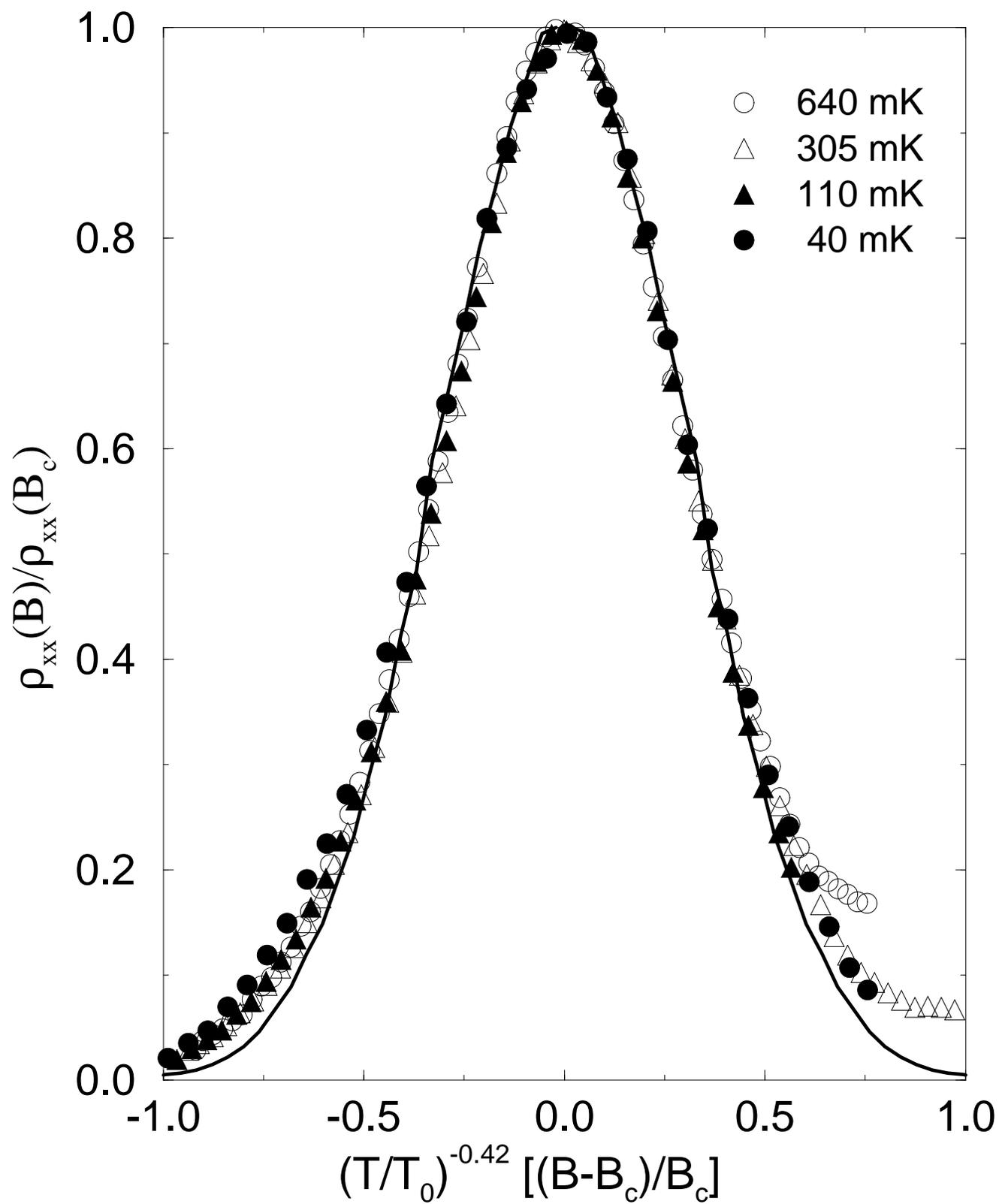
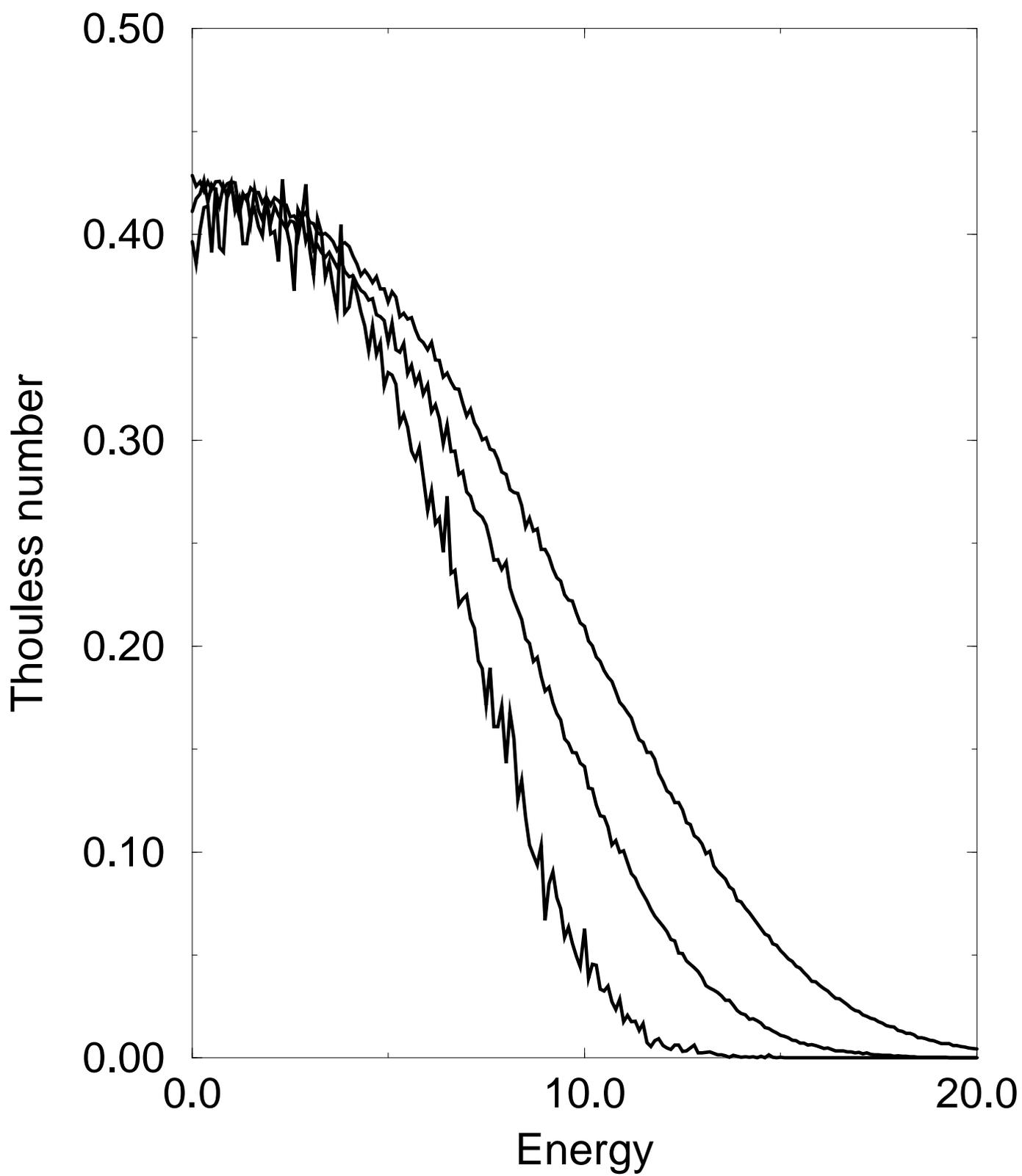
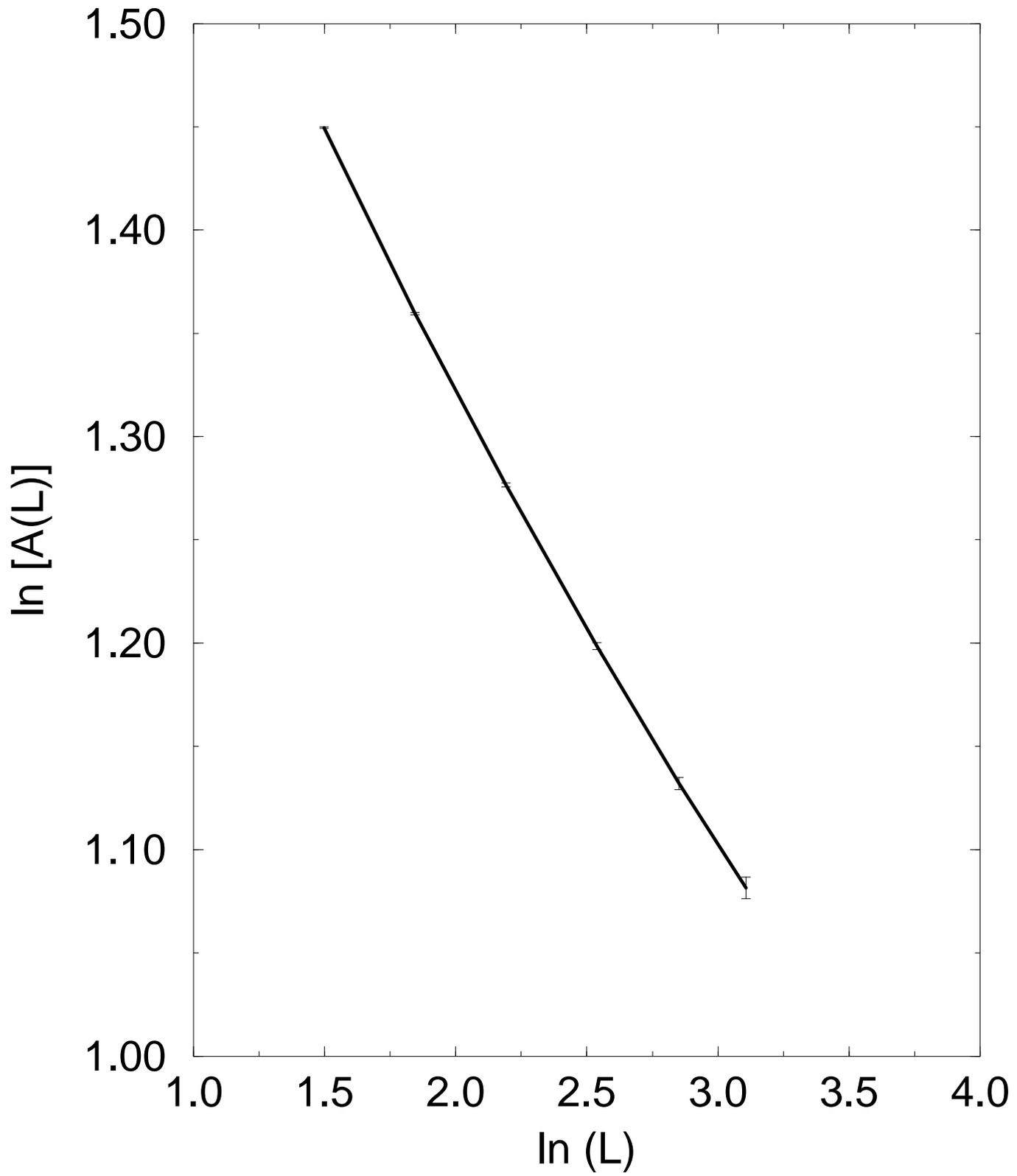
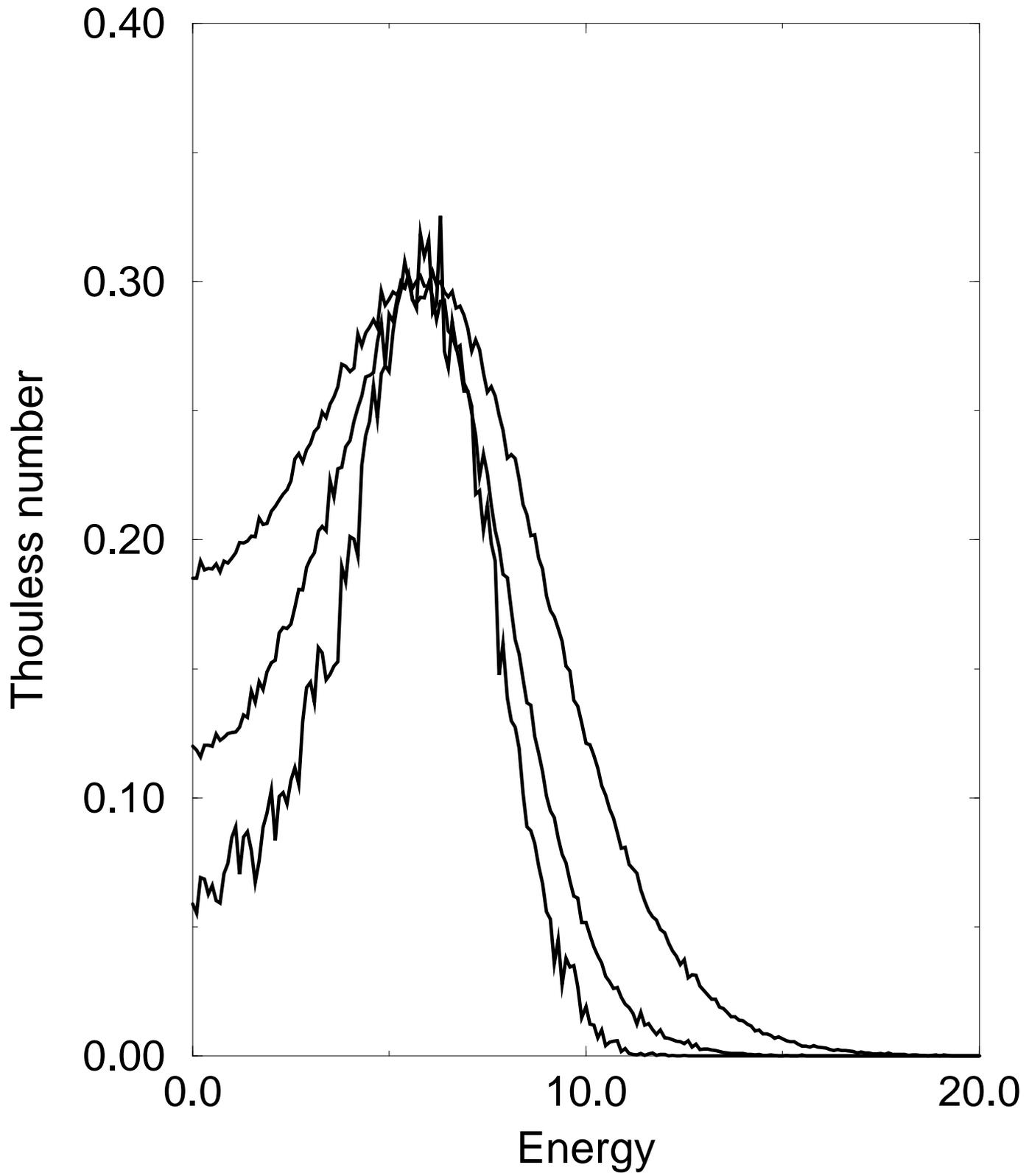
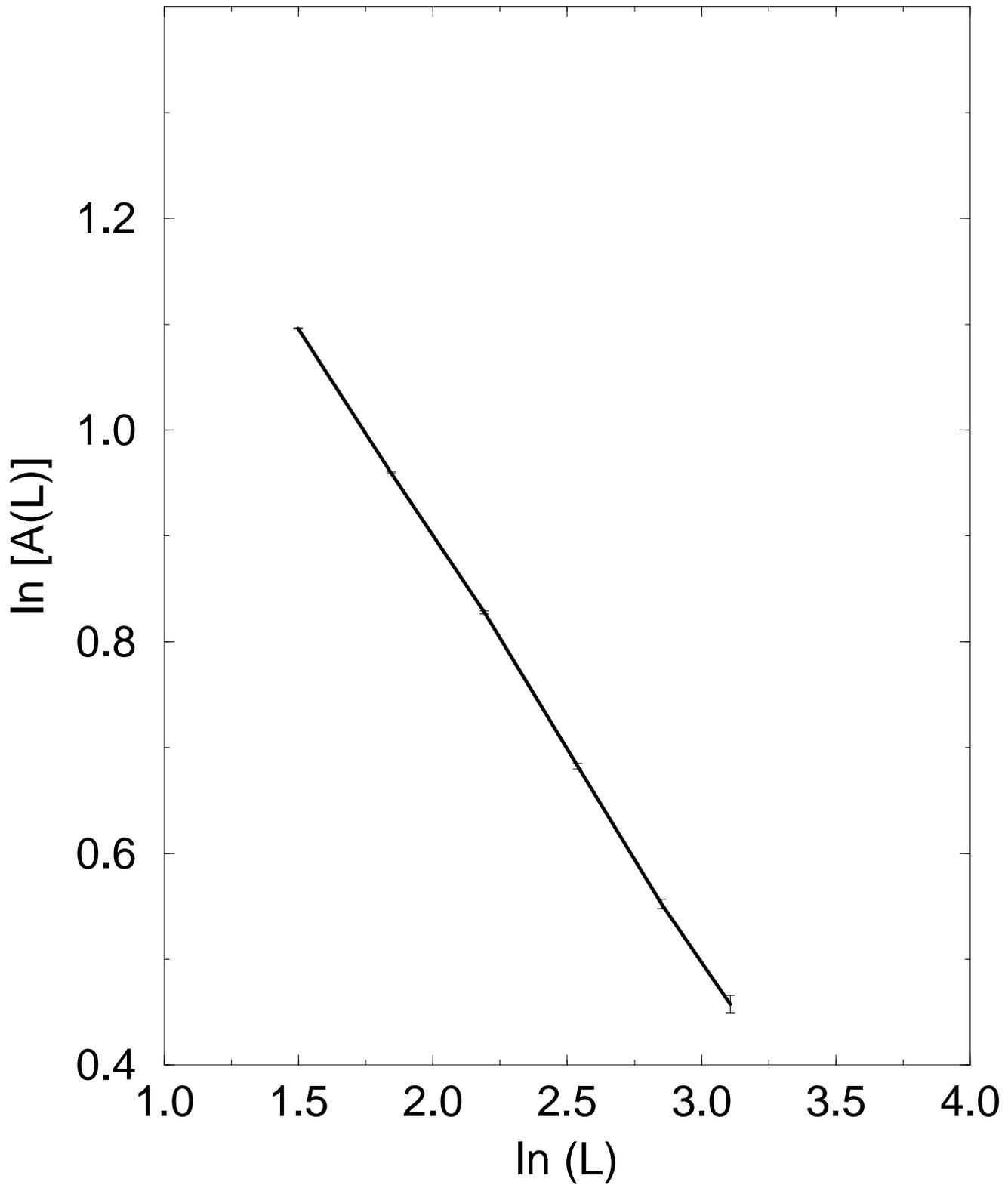
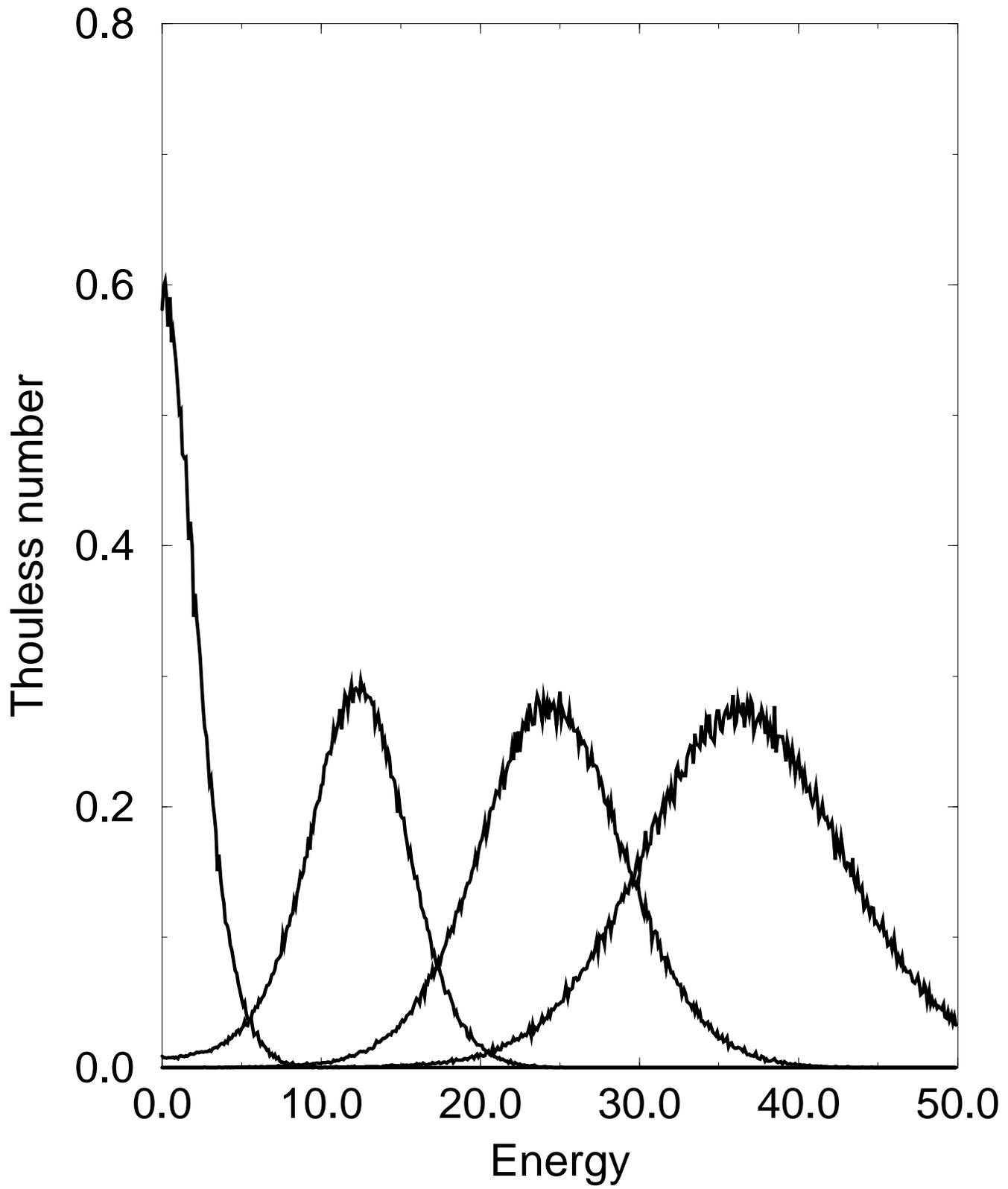
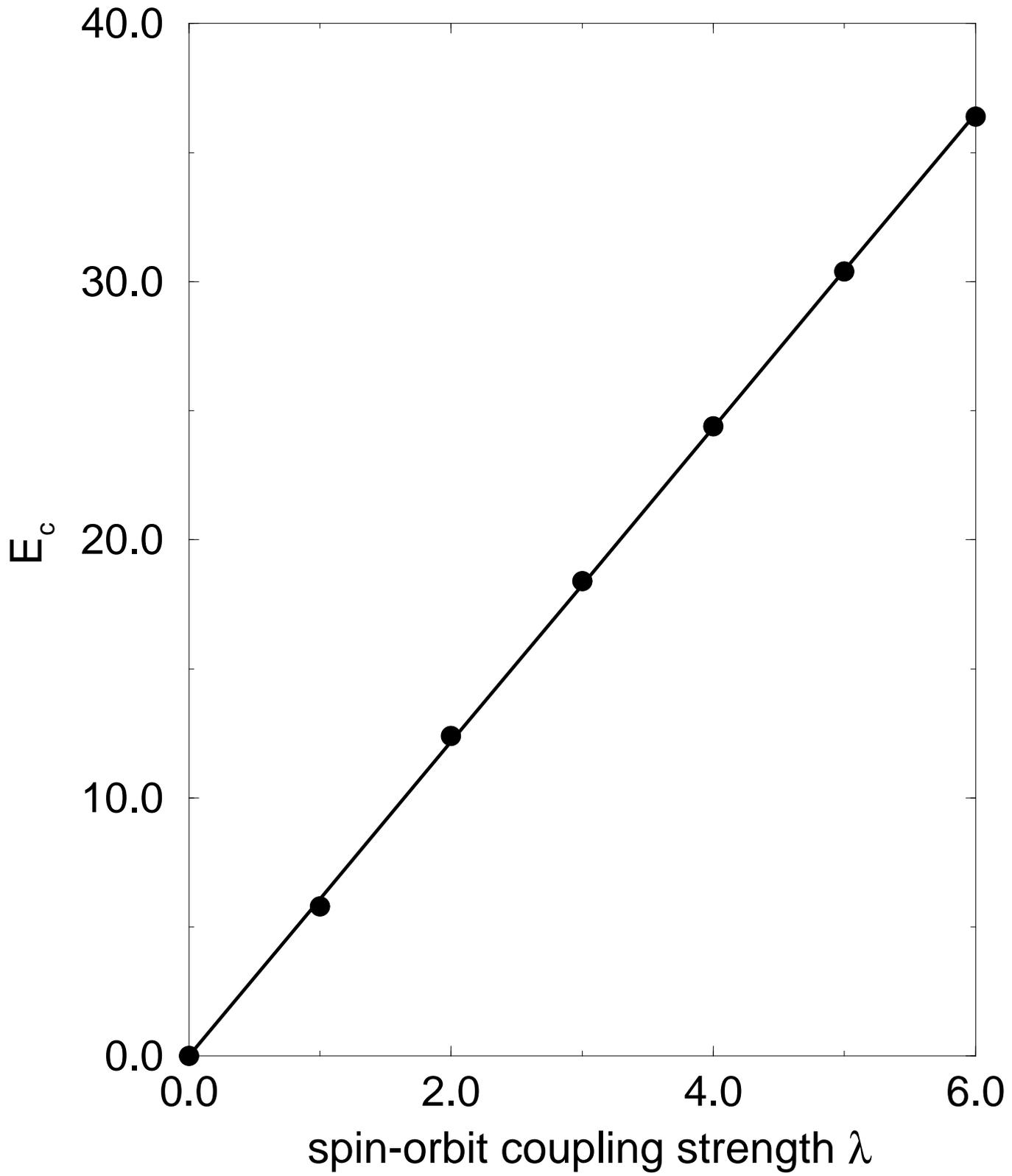
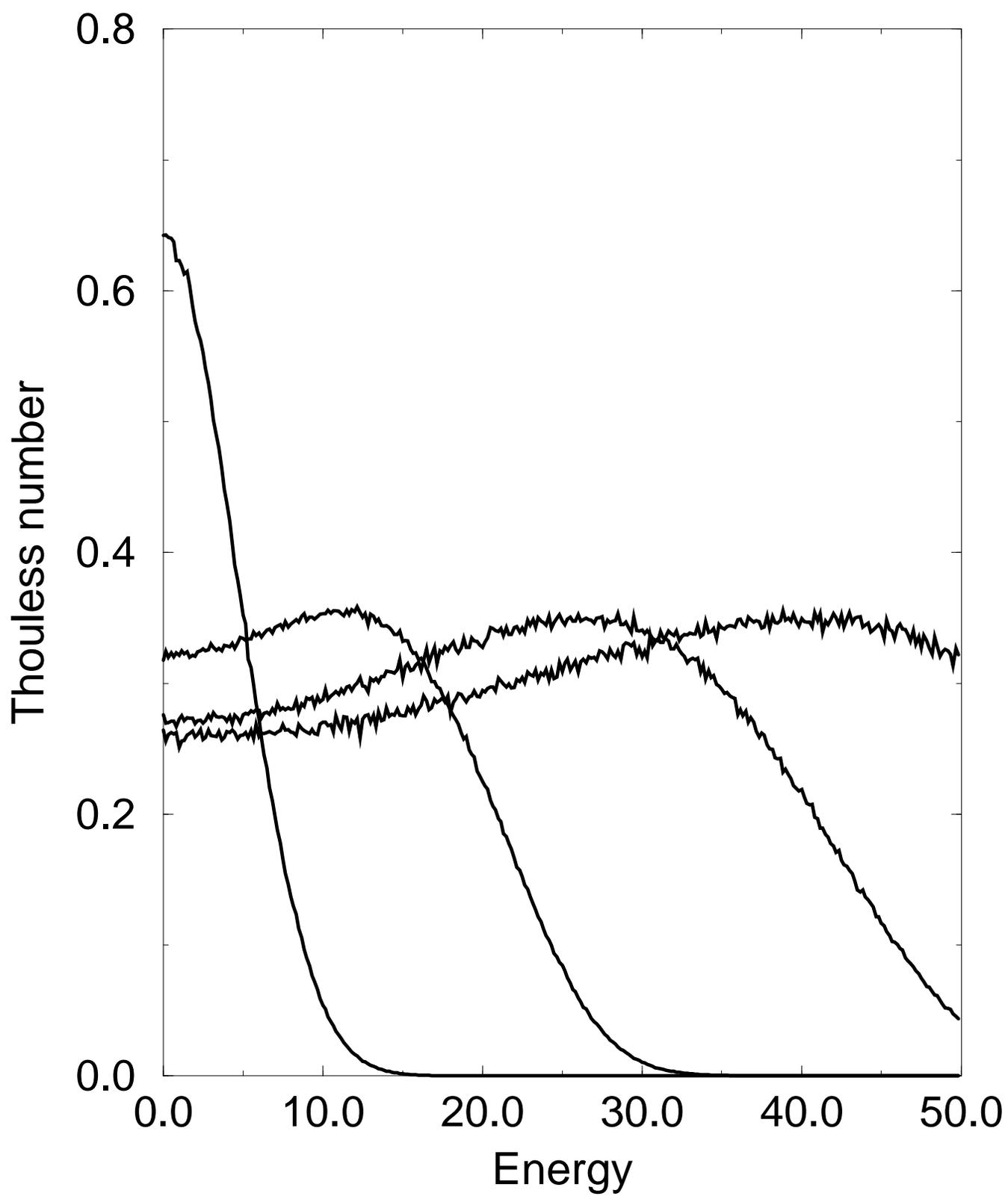
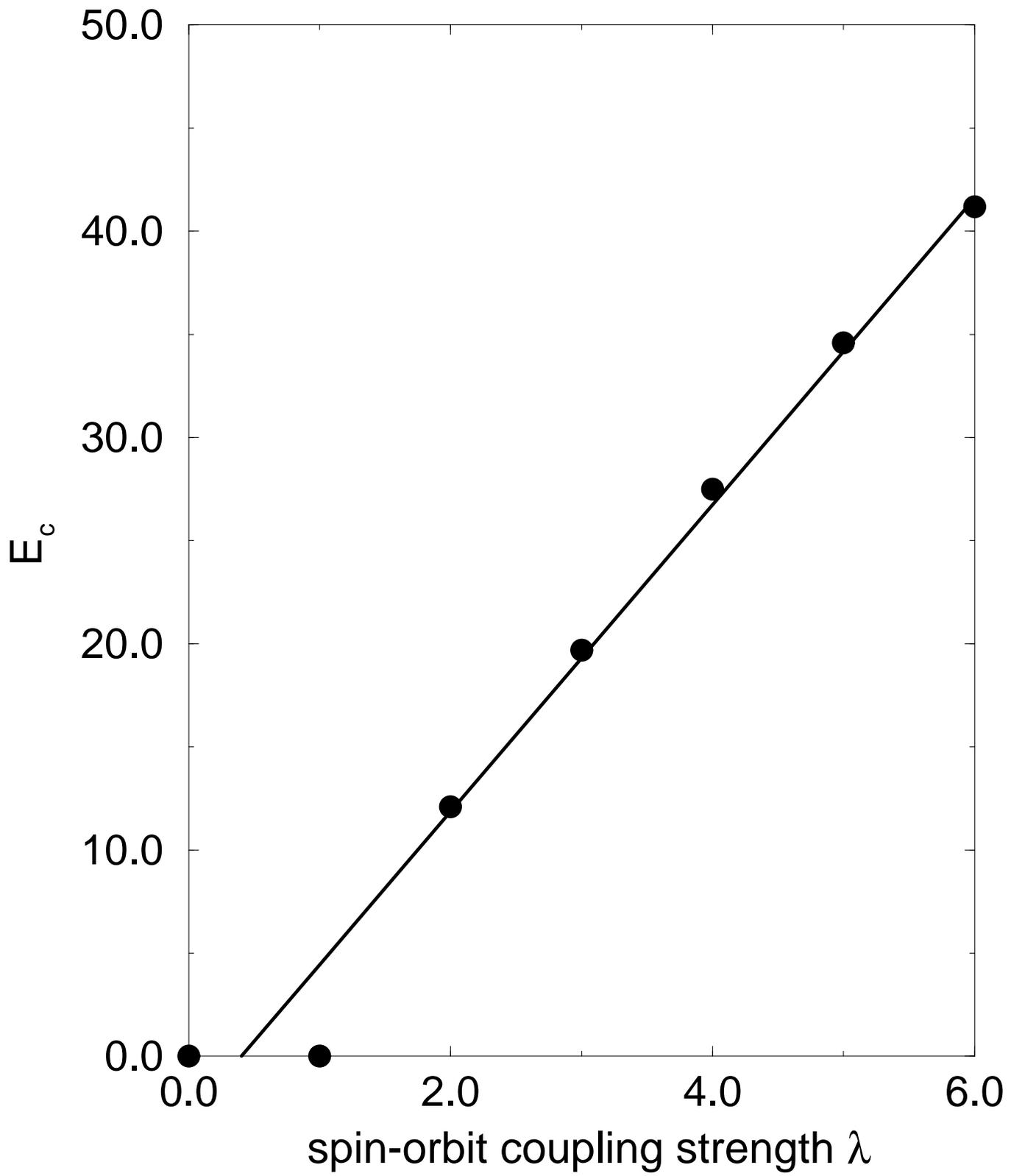
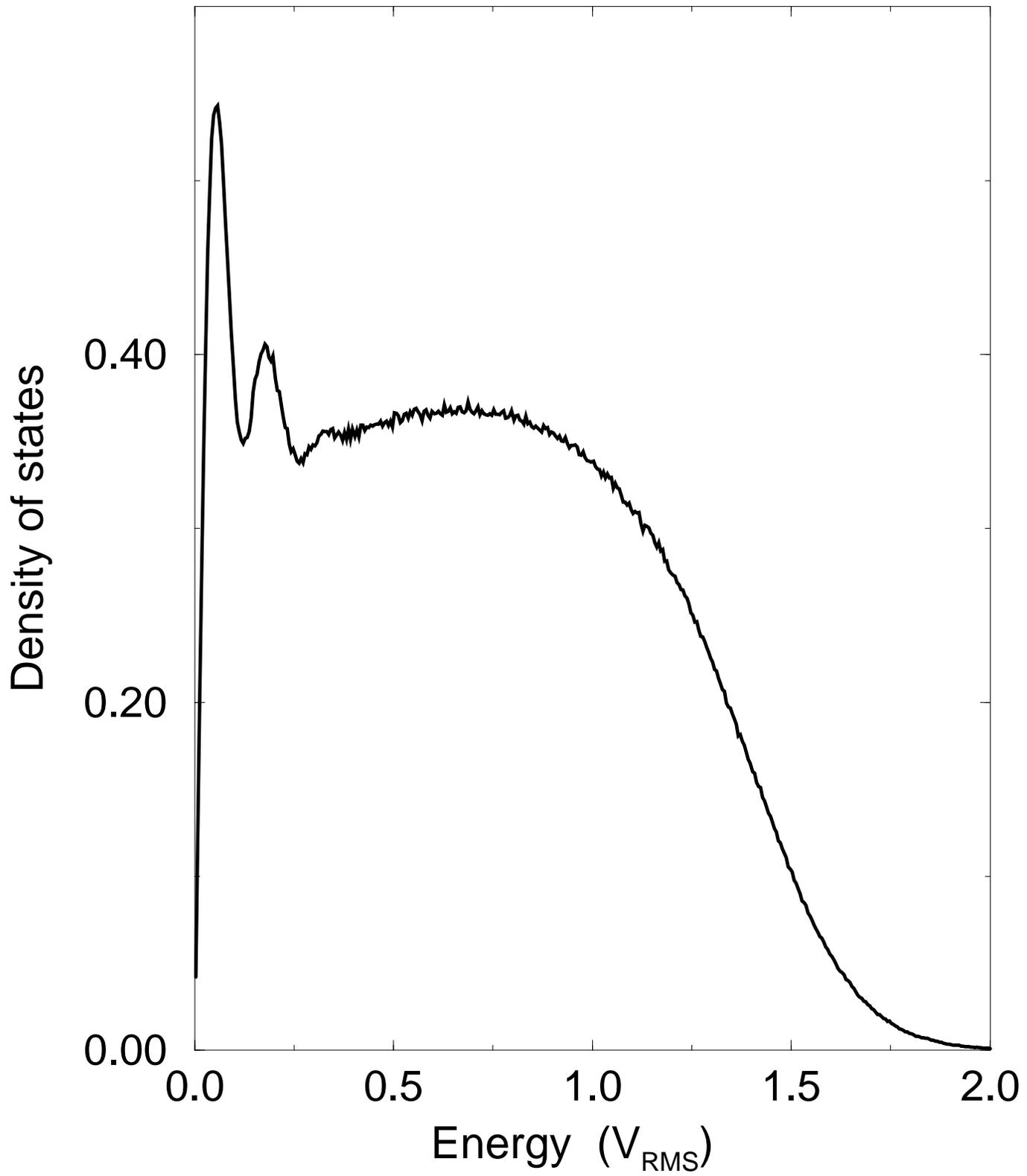
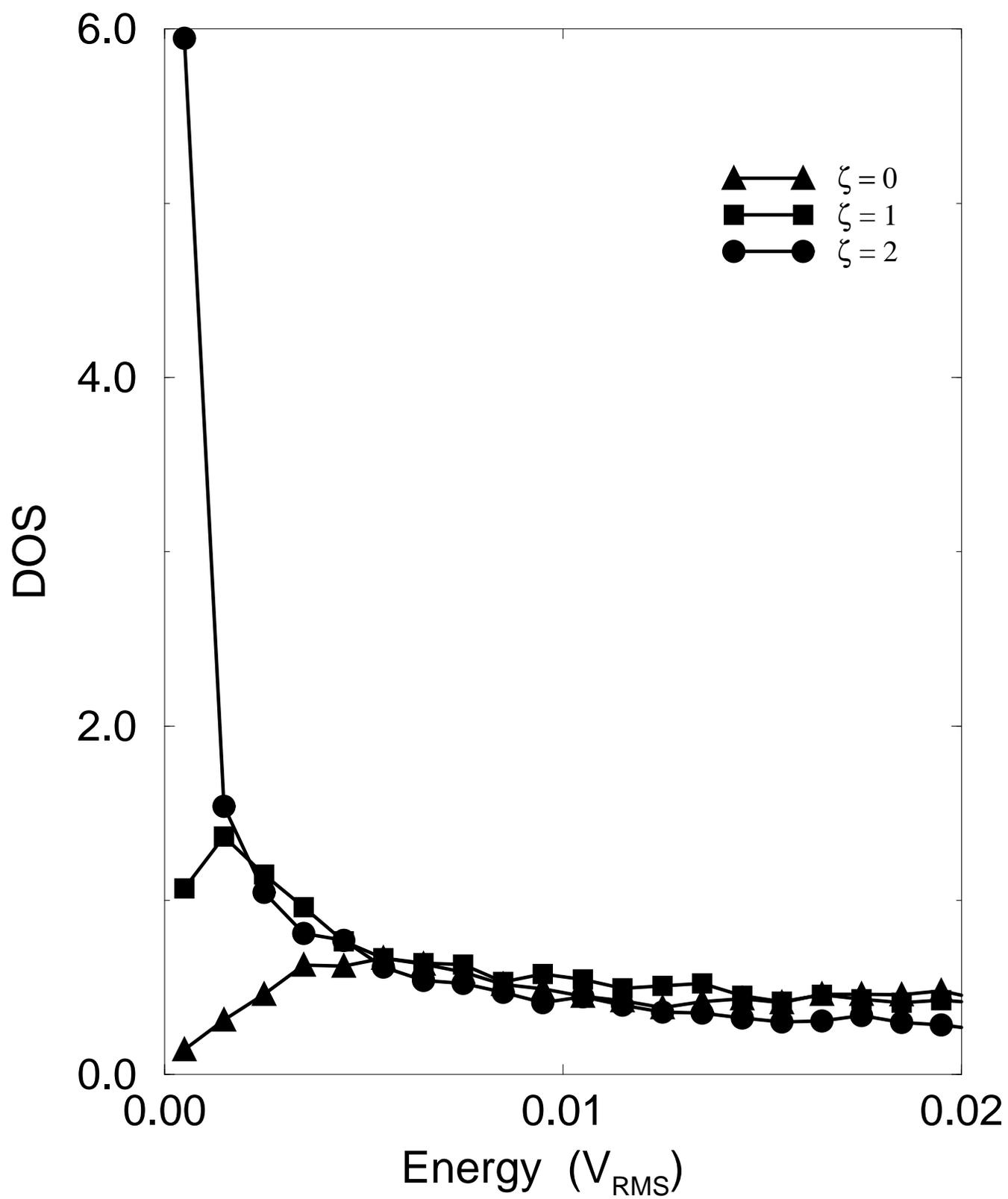
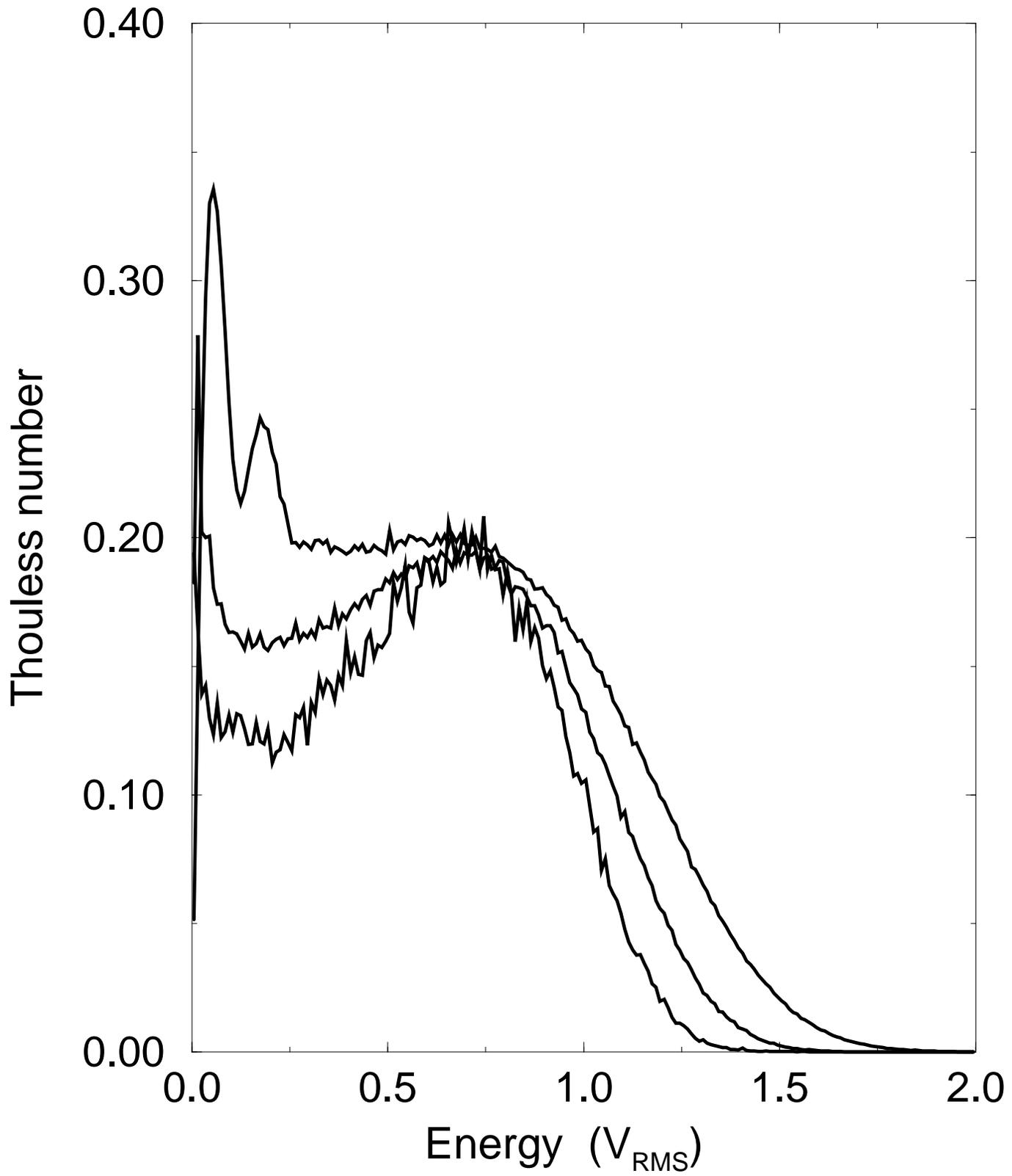
\end{document}